\documentclass[twocolumn]{aastex631}
\usepackage{booktabs}
\usepackage{graphicx}
\usepackage{amsthm,amsmath,amssymb}
\usepackage{mathrsfs}
\usepackage{hyperref}

\begin{document}

\title{Testing General Relativity using Large Scale Structures Photometric Redshift Surveys and Cosmic Microwave Background Lensing Effect}

\author{Shang Li}
\affiliation{School of Physics and Astronomy, Beijing Normal University, Beijing 100875, China}
\affiliation{Institute for Frontiers in Astronomy and Astrophysics, Beijing Normal University, Beijing 100875, China}

\author{Jun-Qing Xia}\thanks{xiajq@bnu.edu.cn}
\affiliation{School of Physics and Astronomy, Beijing Normal University, Beijing 100875, China}
\affiliation{Institute for Frontiers in Astronomy and Astrophysics, Beijing Normal University, Beijing 100875, China}

\begin{abstract}
The $E_G$ statistic provides a valuable tool for evaluating predictions of General Relativity (GR) by probing the relationship between gravitational potential and galaxy clustering on cosmological scales within the observable universe. In this study, we constrain the $E_G$ statistic using photometric redshift data from the Dark Energy Survey (DES) MagLim sample in combination with the Planck 2018 Cosmic Microwave Background (CMB) lensing map. Unlike spectroscopic redshift surveys, photometric redshift measurements are subject to significant redshift uncertainties, making it challenging to constrain the redshift distortion parameter $\beta$ with high precision. We adopt a new definition for this parameter, \(\beta(z) = {f\sigma_8(z)}/{b\sigma_8(z)}\). In this formulation, we reconstruct the growth rate of structure, \(f\sigma_8(z)\), using Artificial Neural Networks (ANN) method, while simultaneously utilizing model-independent constraints on the parameter \(b\sigma_8(z)\), directly obtained from the DES collaboration. After obtaining the angular power spectra \(C_\ell^{gg}\) (galaxy-galaxy) and \(C_\ell^{g\kappa}\) (galaxy-CMB lensing) from the combination of DES photometric data and Planck lensing, we derive new measurements of the $E_G$ statistic: \(E_G = 0.354 \pm 0.146\), \(0.452 \pm 0.092\), \(0.414 \pm 0.069\), and \(0.296 \pm 0.069\) (68\% C.L.) across four redshift bins: \(z = 0.30\), \(0.47\), \(0.63\), and \(0.80\), respectively, which are consistent with the predictions of the standard \(\Lambda\)CDM model. Finally, we forecast the $E_G$ statistic using future photometric redshift data from the China Space Station Telescope, combined with lensing measurements from the CMB-S4 project, indicating an achievable constraint on $E_G$ of approximately 1\%, improving the precision of tests for GR on cosmological scales.
\end{abstract}

\keywords{\href{http://astrothesaurus.org/uat/902}{Large-scale structure of the universe (902)}; \href{http://astrothesaurus.org/uat/1797}{Weak gravitational lensing (1797)}; \href{http://astrothesaurus.org/uat/322}{Cosmic microwave background radiation (322)}; \href{http://astrothesaurus.org/uat/1146}{Observational cosmology (1146)}}

\section{Introduction} \label{sec:intro}

Since the discovery of the accelerated expansion of the universe through supernova observations, many cosmological and gravitational theories have emerged to explain this phenomenon. One of the most well-known models is the \(\Lambda\) cold dark matter (\(\Lambda\)CDM) model, often called the standard model of cosmology. This model is popular because of its simplicity and its success in explaining various types of data, including observations of the Cosmic Microwave Background (CMB) and galaxy surveys.

The \(\Lambda\)CDM model assumes that General Relativity (GR) is the correct theory of gravity. To account for the accelerated expansion of the universe, it introduces a constant called \(\Lambda\), which represents dark energy. This constant creates a type of negative pressure that drives the universe's expansion to speed up. Interestingly, this constant is mathematically identical to what is expected from quantum vacuum energy, but the problem is that its value, which corresponds to an energy scale of about \(10^{-3}\) eV, is vastly different from the predictions made by particle physics. This discrepancy is one of the biggest mysteries in cosmology.

On the other hand, alternative theories known as modified gravity (MG) models have been developed. These models offer different explanations for the universe's expansion. Some of them can reproduce the same expansion history as the \(\Lambda\)CDM model without the need for dark energy, which leads to questioning whether GR is the correct theory of gravity at very large (cosmological) scales.

Although GR and MG models may predict similar rates of expansion when considering the universe as a whole (the "background level"), they usually show differences when we study the small variations, or perturbations, in the universe. These differences can be important for understanding the behavior of cosmic structures, such as galaxies and galaxy clusters, which can help distinguish between these competing theories.

\citet{zhang2007probing} introduced an elegant statistic, $E_G$, designed to distinguish between the \(\Lambda\)CDM+GR framework and MG theories. The $E_G$ statistic is defined as the ratio of the Laplacian of the difference between the two scalar potentials, \(\nabla^2 (\Psi + \Phi)\), to the peculiar velocity divergence field, \(\theta\). In practical applications, \(\nabla^2 (\Psi + \Phi)\) is typically obtained from the cross-correlation between gravitational lensing and galaxy clustering, while the peculiar velocity field is derived from the galaxy-velocity cross-correlation or, equivalently, from the product of the galaxy auto-correlation and the redshift-space distortion (RSD) parameter \(\beta = f/b\), where \(f\) is the growth rate of structure and \(b\) is the galaxy bias.

Importantly, $E_G$ exhibits scale-dependent behavior under modified gravity models, while it remains scale-independent within the \(\Lambda\)CDM framework, as demonstrated by \citet{zhang2007probing} and \citet{pullen2015probing}. This scale-dependence makes $E_G$ a powerful tool for directly testing General Relativity. Moreover, a key advantage of the $E_G$ statistic is that it does not depend on galaxy bias or the amplitude of matter perturbations, providing a more robust probe of gravity on cosmological scales.

The first measurement of the $E_G$ statistic was conducted by \citet{reyes2010confirmation}, using weak lensing measurements of background galaxies as tracers of gravitational lensing. They obtained \(E_G = 0.39 \pm 0.06\) at \(z = 0.32\), confirming the predictions of the \(\Lambda\)CDM model on scales ranging from 10 to 50 \(h^{-1} \text{Mpc}\). Building on this approach, subsequent measurements of $E_G$ have been extended to larger scales, approximately \(70 \, h^{-1} \text{Mpc}\), and over redshifts in the range \(0.2 \leq z \leq 0.6\) \citep{blake2016rcslens,alam2017testing,de2017vimos,amon2018kids+,blake2020testing}.

In addition, \citet{pullen2015probing} proposed using Cosmic Microwave Background (CMB) lensing as an alternative probe for gravitational lensing, enabling $E_G$ to be measured at earlier cosmic times and on larger scales. CMB lensing presents several advantages over galaxy-galaxy lensing. In galaxy-galaxy lensing, source galaxies are typically assigned photometric redshifts, which suffer from significant uncertainties, and are influenced only by foreground galaxies at lower redshifts. In contrast, CMB lensing is sourced at \(z = 1100\), which is far enough to be unaffected by galaxy positions at redshifts \(z \sim 1\). Moreover, CMB lensing is largely free from complex astrophysical effects, such as intrinsic alignments that can bias galaxy lensing measurements, due to the nearly Gaussian intrinsic distribution of CMB photons. This makes CMB lensing a more robust and reliable probe for large-scale gravitational lensing effects.

In a recent study, \citet{zhang2021testing} estimated the $E_G$ statistic at an effective redshift of \(z \sim 1.5\) over scales of 19 to 190 \(h^{-1} \text{Mpc}\), using the Planck 2018 CMB lensing convergence map \citep{aghanim2020planck} and the SDSS eBOSS DR16 quasar clustering catalogs \citep{lyke2020sloan}. The results of this analysis were consistent with the predictions of the \(\Lambda\)CDM+GR model within 1\(\sigma\) significance. 

To estimate $E_G$, in addition to gravitational lensing measurements, it is necessary to measure the growth of structure, which is typically achieved through 3D clustering analysis. This process requires accurate redshift information for tracers, making spectroscopic redshift surveys the preferred method. For photometric redshift (photo-z) surveys, which suffer from lower redshift accuracy, \citet{giannantonio2016cmb} proposed an alternative statistic, $D_G$, which does not require direct measurements of structure growth. However, $D_G$ requires external information on the galaxy bias, which prevents it from directly distinguishing between General Relativity and modified gravity theories, thereby limiting its utility as a probe for testing gravity at cosmological scales.\citep{omori2019dark,marques2020tomographic}

In this work, we aim to estimate the $E_G$ statistic using the magnitude-limited (MagLim) sample from the Dark Energy Survey (DES) \citep{flaugher2015dark} and the Planck 2018 CMB lensing data \citep{aghanim2020planck}. To mitigate the impact of inaccuracies in photometric redshift information on the measurement of structure growth, we calculate the effective redshift of the dataset. We then employ the Artificial Neural Networks (ANN) method to estimate the corresponding structure growth, utilizing prior measurements derived from spectroscopic data. Additionally, we incorporate model-independent constraints on the parameter \(b\sigma_8(z)\), which are directly obtained from the $3\times2$pt analysis of the DES galaxy clustering and lensing probes. This approach helps to reduce the errors introduced by photometric redshift uncertainties, thereby improving the reliability of our $E_G$ estimation.

The structure of this paper is organized as follows: In Section \ref{sec:forma}, we review the theory behind the $E_G$ statistic and the estimator used in our analysis. Section \ref{sec:data} outlines the datasets employed for this study. In Section \ref{sec:result}, we detail the methods used to estimate the RSD parameter, the angular power spectrum and covariance matrix, and present the results obtained from real data. Section \ref{sec:forecast} provides the forecast results for the upcoming galaxy survey, the China Space Station Telescope (CSST). Finally, we summarize our findings and conclusions in Section \ref{sec:summary}.

For the self-consistency test across all datasets, we assume the following cosmological parameters: \(\Omega_m = 0.336\), \(\Omega_b = 0.045\), \(h = 0.670\), \(n_s = 0.959\), and \(\sigma_8 = 0.746\), as given in \citet{abbott2023dark}.

\section{formalism and estimator} \label{sec:forma}

\subsection{\texorpdfstring{$E_G$}{EG} Formalism}

We assume a flat \(\Lambda\)CDM universe described by a perturbed Friedmann-Robertson-Walker (FRW) metric and consider only scalar perturbations. The metric in the conformal Newtonian gauge is expressed as
\begin{equation}
    ds^2 = a(\tau)[-(1+2\Psi)d\tau^2 + (1-2\Phi)dx^2]~,
\end{equation}
where \(\Psi\) and \(\Phi\) represent the weak-field potentials in the time and space metric components, respectively. In GR, when anisotropic stress is negligible, the two potentials are equal, i.e., \(\Phi = \Psi\). However, this equality typically does not hold in MG models, leading to a phenomenon known as gravitational slip.

Following \citet{zhang2007probing}, the $E_G$ statistic in Fourier space can be defined as:
\begin{equation}
E_G(z, k) = \frac{c^2k^2(\Psi + \Phi)}{3H_0^2(1+z)\theta(k)}~,
\label{equ:eg_2007}
\end{equation}
where \(\theta = \nabla \cdot \mathbf{v}/H(z)\) is the divergence of the peculiar velocity field \(\mathbf{v}\), and \(H(z)\) is the Hubble expansion rate at redshift \(z\). In linear perturbation theory, \(\theta(k)\) can be expressed as \(\theta = -f(z)\delta_m(k, z)\), where \(f\) is the linear growth rate, which in GR is given by \(f \approx [\Omega_m(z)]^{0.55}\). Here, \(\Omega_m(z)\) represents the matter density parameter at redshift \(z\).

According to the Poisson equation, \( k^2 \Phi = \frac{3}{2} H_0^2 \Omega_{m0}(1+z) \delta_m \), and assuming the relation \(\Phi = \Psi\), we can simplify the $E_G$ statistic in GR as follows: $E_G(z) = {\Omega_{m0}}/{f(z)}$. Under these assumptions, the $E_G$ statistic becomes independent of scale, meaning that it remains constant at a given redshift in GR. In contrast, in MG models, the $E_G$ statistic typically exhibits scale dependence. This scale dependence in MG models arises from deviations in the relationship between the gravitational potentials \(\Psi\) and \(\Phi\). Consequently, measuring $E_G$ at different scales could distinguishing between GR and alternative theories of gravity.

Following Eq.(\ref{equ:eg_2007}), the $E_G$ statistic can also be expressed in terms of power spectra as:
\begin{equation}
    E_G(k,z)=\frac{c^2 P_{\nabla^2(\Psi+\Phi)g}(k)}{3H_0^2(1+z)P_{\theta g}(k)}~,
\end{equation}
where $P_{\nabla^2(\Psi+\Phi)g}$ is the galaxy-$\nabla^2(\Psi+\Phi)$ cross-power spectrum and $P_{\theta g}$ is the galaxy-peculiar velocity cross-power spectrum. By projecting the 3-D power spectrum onto a 2-D spherical surface, $E_G$ can be estimated as: 
\begin{equation}
    E_G(\ell,\bar{z})=\frac{c^2 C_{\ell}^{\mathrm{g\kappa}}}{3H_0^2 C_{\ell}^{\mathrm{g\theta}}}~,
\end{equation}
where $\bar{z}$ is the effective redshift of the galaxy survey,  $C_{\ell}^{\mathrm{g\kappa}}$ is the measured lensing convergence-galaxy cross-correlation angular power spectrum and $C_{\ell}^{\mathrm{g\theta}}$ is the velocity-galaxy cross-correlation angular power spectrum. 

In practice, measuring $C_{\ell}^{\mathrm{g\theta}}$ directly is challenging. However, we can utilize the relationship $C_{\ell}^{\mathrm{g\theta}} = C_{\ell}^{\mathrm{gg}} \cdot \beta $, where $\beta = f/b$ is the RSD parameter, $b$ is the linear bias parameter of the galaxy survey. This allows us to replace $C_{\ell}^{\mathrm{g\theta}}$ with the product of the galaxy auto-correlation power spectrum $C_{\ell}^{\mathrm{gg}}$ and $\beta$, which are easier to measure. Then, the modified expression for $E_G$ becomes \citep{pullen2015probing}
\begin{equation}
{E_G}(\ell,\bar{z})=\Gamma(\bar{z})\frac{{C}_{\ell}^{\mathrm{g\kappa}}}{\beta(\bar{z}) {C}_{\ell}^{\mathrm{gg}}}~,
\label{eq:EG}
\end{equation}
where $\Gamma(z)$ is the calibration factor.

\subsection{angular power spectrum}

In the Limber approximation, the angular cross correlation power spectrum between two tracers can be described generally as \citep{limber1953analysis}:
\begin{equation}
C_{\ell}^{\mathrm{XY}} = \int dz \, \frac{H(z)}{c} \, \frac{W_{\mathrm{X}}(z) W_{\mathrm{Y}}(z)}{\chi^2(z)} \, P_{\mathrm{XY}}\left(\frac{\ell + 1/2}{\chi(z)}, z\right)~,
\label{eq:clxy}
\end{equation}
where \(\chi(z)\) is the comoving distance at redshift \(z\), \(W_{\mathrm{X/Y}}(z)\) are the window functions for the respective tracers, \(c\) is the speed of light, and \(P_{\mathrm{XY}}\) is the 3D power spectrum of two tracers. In the estimation of the $E_G$ statistic, we mainly focus on the linear regime. Therefore, we can substitute the matter power spectrum \(P_{\mathrm{mm}}(k, z)\) in place of \(P_{\mathrm{XY}}(k, z)\), since the linear matter fluctuations dominate at large scales. The tracer-specific effects are then incorporated into the window functions.

For the galaxy survey, the window function is given by:$W_{\mathrm{g}}(z) = b(z) n(z)$, where \(b(z)\) is the galaxy bias and \(n(z)\) is the redshift distribution of the galaxy sample. The lensing convergence window function for the CMB at redshift $z_*=1100$ is expressed as:
\begin{equation}
W_{\mathrm{\kappa}}(z,z_*)=\frac{3H_0^2\Omega_{m,0}}{2cH(z)}(1+z)W(z,z_*)~,
\label{eq:lensing_window_source}
\end{equation}
where the lensing convergence kernel function $W(z,z_*)$ is given by:
\begin{equation}
W(z,z_*)=\chi(z)\left(1-\frac{\chi(z)}{\chi(z_*)}\right)~,
\end{equation}
with $\chi(z)$ being the comoving distance at redshift $z$, and $\chi(z_*)$ the comoving distance to the CMB. Using these window functions, the angular power spectra $C_{\ell}^{\mathrm{gg}}$ and $C_{\ell}^{\mathrm{g\kappa}}$ can be described as 
\begin{equation}
    C_{\ell}^{\mathrm{gg}}= \int d z \frac{H(z)}{c} \left[ \frac{b(z)n(z)}{\chi(z)}\right]^2 P_{\mathrm{mm}}\left(\frac{\ell+1 / 2}{\chi(z)}, z\right)~,
\label{eq:clgg}
\end{equation}
and
\begin{equation}
\begin{split}
    C_{\ell}^{\mathrm{g\kappa}}&= \frac{3H_0^2\Omega_{m,0}}{2c^2} \int d z (1+z) \frac{W(z,z_*)}{\chi^2(z)}\\
   &\times b(z)n(z) P_{\mathrm{mm}}\left(\frac{\ell+1 / 2}{\chi(z)}, z\right)~.
\end{split}
\label{eq:clgk}
\end{equation}
These expressions provide the theoretical predictions for the galaxy auto-correlation power spectrum and the galaxy-CMB lensing cross-correlation power spectrum.

\subsection{calibration factor}
In addition to the angular power spectra \(C_{\ell}^{\mathrm{gg}}\) and \(C_{\ell}^{\mathrm{g\kappa}}\), the calibration factor \(\Gamma(z)\) is needed for estimating \(E_G\) in Eq.(\ref{eq:EG}). Following \citet{pullen2016constraining} and \citet{yang2018calibrating}, in the calibration factor \(\Gamma(z)\) it is essential to consider several additional factors beyond the standard normalization term to ensure accuracy:
\begin{equation}
\Gamma(\ell,z) = C_{\Gamma}C_b \frac{2c}{3H_0} \left[ \frac{H(z)n(z)}{H_0(1+z)W(z,z_*)} \right],\label{eq:calibration_factor}
\end{equation}
where \(C_{\Gamma}\) and \(C_b\) are extra calibration factors that account for the broad redshift distribution and the lensing kernel, as well as for scale-dependent bias due to nonlinear clustering. These are expressed as:
\begin{equation}
C_{\Gamma}(\ell,z) = \frac{W(z,z_*)(1+z)}{2n(z)}\frac{c}{H(z)}\frac{C_{\ell}^{\mathrm{mg}}}{Q_{\ell}^{\mathrm{mg}}},
\end{equation}
and
\begin{equation}
C_b(\ell,z) = \frac{C_{\ell}^{\mathrm{gg}}}{b(\bar{z}) C_{\ell}^{\mathrm{mg}}},\label{eq:extra_cab_factor}
\end{equation}
where \(C_{\ell}^{\mathrm{mg}}\) is the angular power spectrum that combines galaxy bias with the matter power spectrum:
\begin{equation}
C_{\ell}^{\mathrm{mg}} = \int_{0}^{\infty} dz \frac{H(z)}{c} b(z) n^2(z) \chi^{-2}(z) P_{\mathrm{mm}} \left( \frac{\ell + 1/2}{\chi(z)}, z \right),
\end{equation}
and \(Q_{\ell}^{\mathrm{mg}}\) is given by:
\begin{equation}
\begin{split}
Q_{\ell}^{\mathrm{mg}} &= \frac{1}{2}\int_{0}^{\infty} d z W(z,z_*)b(z)n(z)\chi^{-2}(z)\\
&\times (1+z)P_{\mathrm{mm}}\left(\frac{\ell+1 / 2}{\chi(z)}, z\right)
\end{split}
\label{eq:cql_mg}
\end{equation}
These equations provide the framework to compute the necessary calibration factors \(\Gamma(z)\), which are crucial for correcting the estimation of \(E_G\). The calibration accounts for both the impact of galaxy redshift distribution and the effects of nonlinear clustering on bias and lensing.

\section{Observational Data} \label{sec:data}
\subsection{DES magnitude-limited sample (MagLim)}\label{des}
The DES \citep{flaugher2015dark} is a large-scale imaging survey designed to cover approximately 5000 square degrees of the southern sky, employing five broadband filters (\(grizY\)) to observe galaxies. DES operates with a 570-megapixel camera mounted on the 4-meter Blanco telescope at the Cerro Tololo Inter-American Observatory (CTIO) in Chile. One of the key scientific goals of DES is to provide stringent constraints on cosmological parameters, including the dark energy equation of state parameter \(w\).

This work utilizes data from the first three years (Y3) of DES observations, collected between August 2013 and February 2016. The analysis focuses on the MagLim galaxy sample derived from the Y3 GOLD catalog, applying the same selection criteria as \citet{rodriguez2022dark}. The MagLim sample is defined by an $i$-band magnitude cut that depends linearly on the photometric redshift, facilitating the inclusion of a greater number of higher-redshift galaxies. The photometric redshifts for the MagLim sample are estimated using the Directional Neighborhood Fitting (DNF) algorithm \citep{de2016dnf}.

\begin{figure}[t]
    \centering
    \includegraphics[width=0.45\textwidth]{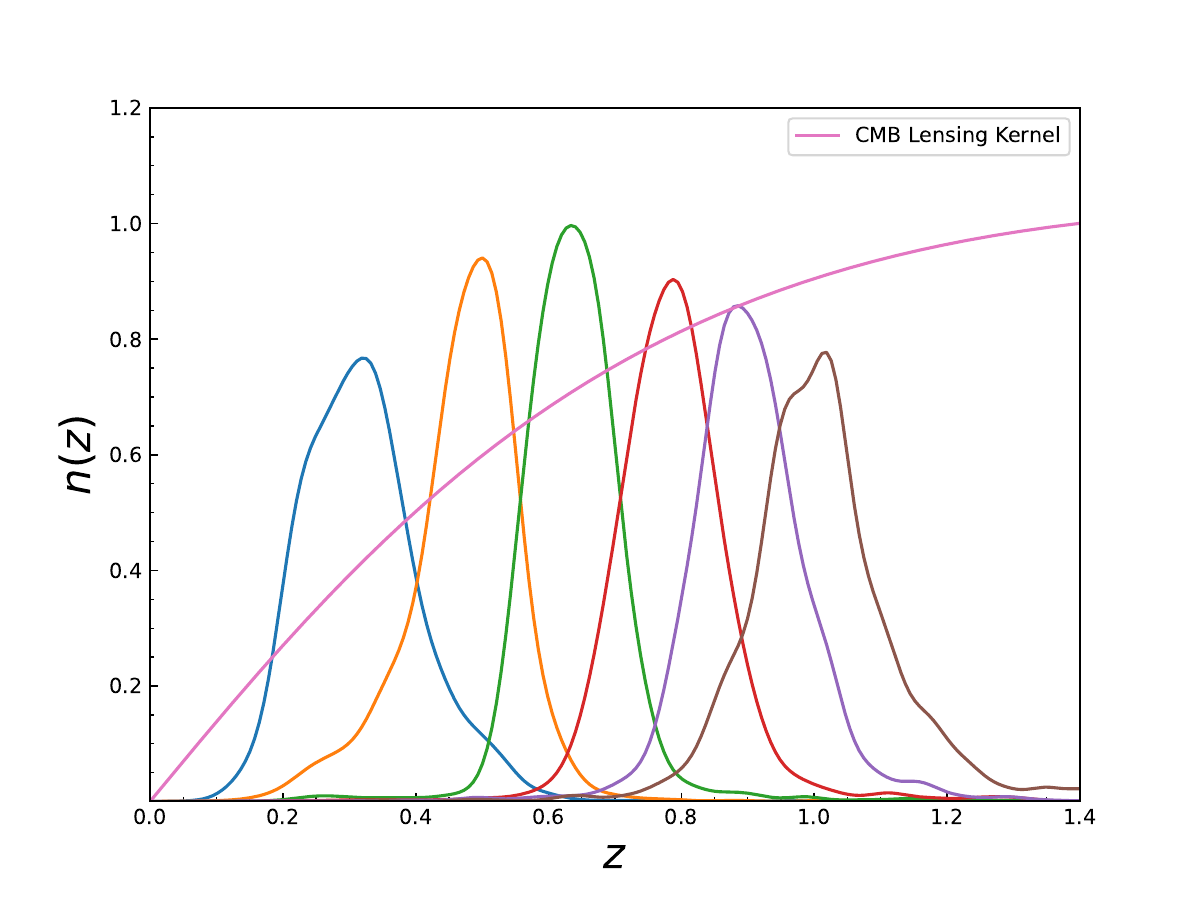}
  \caption{Redshift distributions of the DES Y3 MagLim sample. The kernel function of the CMB lensing is also shown in the figure. For clarity, all results have been normalized.}
  \label{fig:dndz}
\end{figure}

In line with the method used by \citet{porredon2022dark} and \citet{sanchez2022mapping}, we divide the MagLim galaxies into six tomographic redshift bins ranging from \(z = 0.2\) to \(z = 1.05\), with bin edges at \( [0.20, 0.40, 0.55, 0.70, 0.85, 0.95, 1.05] \). Figure \ref{fig:dndz} shows the redshift distribution for each tomographic bin. However, consistent with the findings of \citet{abbott2022dark}, we adopt a conservative approach and exclude the two highest redshift bins from our analysis. This decision stems from issues identified after unblinding the data, where significant discrepancies were observed in both the clustering and galaxy-galaxy lensing signals for galaxies at \(z > 0.85\), leading to poor fits with the cosmological models under consideration.

The MagLim catalog assigns a weight to each galaxy to correct for observational systematics, as described in \citet{rodriguez2022dark}. Using these weights, we construct a map of the number of sources per pixel, where the total number of galaxies in pixel \(p\) is computed as \( N_p = \sum_{i \in p} w_i \), with \( w_i \) being the weight of the \(i\)-th galaxy. For the pixelization scheme, we choose the HEALPix resolution parameter \( N_{\text{side}} = 2048 \), which matches the resolution of the CMB lensing mask map discussed in Section \ref{subsec:planck_len}.

Each pixel in the galaxy overdensity map is then defined by the relation \citep{marques2024cosmological}:
\begin{equation}
    \delta_p = \left( \frac{1}{f_p} \frac{N_p}{\bar{n}} \right) - 1~,
\end{equation}
where \( \bar{n} \) represents the mean number of sources in the unmasked pixels and is calculated as:
\begin{equation}
    \bar{n} = \frac{\sum_p N_p}{\sum_p f_p}~.
\end{equation}
Here, \( f_p \) denotes the fractional coverage of each pixel, which accounts for the DES mask. The DES mask is provided at a higher resolution of \( N_{\text{side}} = 4096 \), with values of \( f_p \) ranging from 0.8 to 1 for effectively observed regions. To match the galaxy map resolution, we average the \( f_p \) values from the higher-resolution pixels, degrading the mask map to \( N_{\text{side}} = 2048 \).

Following this processing, we obtain the galaxy density map and the corresponding mask map, which together cover an area of 4143 square degrees of the sky. These maps will be used for the galaxy clustering analysis in this work.

\subsection{CMB Lensing}\label{subsec:planck_len}
Gravitational lensing of the Cosmic Microwave Background (CMB) can be detected because of the detailed understanding of the primordial CMB's statistical properties. As CMB photons travel from the last scattering surface to Earth, their paths are deflected by intervening matter, causing subtle distortions in the observed anisotropies. These distortions alter the statistical characteristics of the CMB, enabling the reconstruction of a map of the gravitational potential responsible for the deflection. This gravitational potential provides valuable insights into the growth and distribution of cosmic structures.

Since the CMB lensing map traces the matter distribution directly, it acts as an unbiased tracer of the Universe’s matter density field. For this analysis, we use the minimum-variance (MV) estimate of the gravitational lensing convergence, as reconstructed from the CMB temperature and polarization measurements in the Planck 2018 data release \citep{aghanim2020planck}.

Specifically, we utilize the \texttt{COM\_Lensing\_4096\_R3.00} dataset, which is based on both temperature and polarization measurements. This dataset comes with a survey mask covering approximately 70\% of the sky at \( N_{\text{side}} = 2048 \) and includes the noise estimate for the Planck lensing reconstruction, \( N_{\ell}^{\kappa \kappa} \). The MV lensing potential estimates are derived from SMICA DX12 CMB maps, with the lensing convergence available in spherical harmonics coefficients, \( a_{\ell m} \), up to \( \ell_{\text{max}} = 4096 \).

For our analysis, we limit the range of multipoles to \( 8 \leq \ell \leq 2048 \), excluding higher multipoles due to the significant reconstruction noise at small scales. This choice focuses on quasi-linear scales, where the contribution from non-linear effects is minimal relative to statistical errors. Moreover, we adopt a conservative multipole range of \( \ell_{\text{min}} = 80 \) and \( \ell_{\text{max}} = 400 \), as we will discuss in more detail in Section \ref{sec:result}, to ensure that non-linear scales do not dominate our results.

\section{Numerical results}\label{sec:result}
\subsection{RSD parameter}

In cosmological analyses, the growth of structure is typically measured through 3D clustering, which necessitates precise redshift information for tracers. As such, spectroscopic redshift surveys are generally preferred due to their superior accuracy, whereas photometric redshift surveys, such as DES, suffer from lower redshift precision. Consequently, in our analysis, we do not directly estimate the RSD parameter \(\beta = f/b\) from photometric redshift data. Instead, we adopt a modified expression, \(\beta(z) = \hat{f}/\hat{b}\), where \(\hat{f} = f\sigma_8(z)\) and \(\hat{b} = b\sigma_8(z)\) \citep{wenzl2024constraining}. By leveraging the DES galaxy and lensing probes, we can estimate \(\hat{b}\), while \(\hat{f}\) is derived from previous spectroscopic measurements of the linear growth rate, employing a regression algorithm.

\begin{table*}[t]
  \caption{The $\hat{f}$ data compilation used in the analysis. References:(1)a:SDSS-LRG,b:VVDS,c:2dFGRS \citep{Song:2008qt};(2)2MRS \citep{Davis:2010sw};(3)a:2MRS,b:SnIa+IRAS \citep{Hudson:2012gt};(4)SnIa+IRAS \cite{Turnbull:2011ty};(5)SDSS-LRG-200 \citep{Samushia:2011cs};(6)WiggleZ \citep{Blake:2012pj};(7)6dFGS \citep{Beutler:2012px};(8)SDSS-BOSS \citep{Tojeiro:2012rp};(9)VIPERS \citep{delaTorre:2013rpa};(10)SDSS-DR7-LRG \citep{Chuang:2012qt};(11)GAMA \citep{Blake:2013nif};(12)a:BOSS-LOWZ,b:SDSS DR10 and DR11 \citep{Sanchez:2013tga};(13)SDSS-MGS \citep{Howlett:2014opa};(14)SDSS-veloc \citep{Feix:2015dla};(15)FastSound \citep{Okumura:2015lvp};(16)SDSS-CMASS \citep{Chuang:2013wga};(17)BOSS DR12 \citep{Alam:2016hwk};(18)BOSS DR12 \citep{Beutler:2016arn};(19)VIPERS v7 \citep{Wilson:2016ggz};(20)BOSS LOWZ \citep{Gil-Marin:2016wya};(21)VIPERS \citep{Hawken:2016qcy};(22)6dFGS+SnIa \citep{Huterer:2016uyq};(23)VIPERS PDR2 \citep{Pezzotta:2016gbo};(24)SDSS DR13 \citep{Feix:2016qhh};(25)2MTF \citep{Howlett:2017asq};(26)VIPERS PDR2 \citep{Mohammad:2017lzz};(27)BOSS DR12 \citep{Wang:2017wia};(28)SDSS DR7 \citep{Shi:2017qpr};(29)SDSS-IV \citep{Gil-Marin:2018cgo};(30)SDSS-IV \citep{Hou:2018yny};(31)SDSS-IV \citep{Zhao:2018jxv};(32)VIPERS PDR2 \citep{Mohammad:2018mdy};(33)BOSS DR12 voids \citep{Nadathur:2019mct};(34)2MTF 6dFGSv \citep{Qin:2019axr};(35)SDSS-IV \citep{Icaza-Lizaola:2019zgk}}
  \label{tab:data-rsd}
  \begin{tabular}{ccc||ccc||ccc}
    \hline
    $z$ & $\hat{f}$ & Refs.& $z$ & $\hat{f}$ & Refs.& $z$ & $\hat{f}$ & Refs.\\
    \hline
    $0.35$ & $0.440\pm 0.050$ & [1a]  & $0.77$ & $0.490\pm 0.18$ & [1b] & $0.17$ & $0.510\pm 0.060$ & [1c]\\  
    $0.02$& $0.314 \pm 0.048$ &  [2][3a] & $0.02$& $0.398 \pm 0.065$ & [3b][4] & $0.25$ & $0.3512\pm 0.0583$ & [5]    \\  
    $0.37$ & $0.4602\pm 0.0378$ & [5]  &$0.25$ & $0.3665\pm0.0601$ & [5] & $0.37$ & $0.4031\pm0.0586$ & [5] \\ 
    $0.44$ & $0.413\pm 0.080$ & [6] &   $0.60$ & $0.390\pm 0.063$ & [6]  &  $0.73$ & $0.437\pm 0.072$ & [6]  \\
    $0.067$ & $0.423\pm 0.055$ & [7] &   $0.30$ & $0.407\pm 0.055$ & [8] &  $0.40$ & $0.419\pm 0.041$ & [8] \\ 
    $0.50$ & $0.427\pm 0.043$ & [8]  & $0.60$ & $0.433\pm 0.067$ & [8] &   $0.80$ & $0.470\pm 0.080$ & [9]  \\ 
    $0.35$ & $0.429\pm 0.089$ & [10] &  $0.18$ & $0.360\pm 0.090$ & [11] &  $0.38$ & $0.440\pm 0.060$ & [11]\\
    $0.32$ & $0.384\pm 0.095$ & [12a] & $0.32$ & $0.48 \pm 0.10$ & [12b] &  $0.57$ & $0.417 \pm 0.045$ & [12b]  \\
    $0.15$ & $0.490\pm0.145$ & [13]  &  $0.10$ & $0.370\pm 0.130$ & [14] & $1.40$ & $0.482\pm 0.116$ & [15]\\
    $0.59$ & $0.488\pm 0.060$ & [16] & $0.38$ & $0.497\pm 0.045$ & [17]  & $0.51$ & $0.458\pm 0.038$ & [17] \\
    $0.61$ & $0.436\pm 0.034$ & [17] &  $0.38$ & $0.477 \pm 0.051$ & [18] & $0.51$ & $0.453 \pm 0.050$ & [18]\\
    $0.61$ & $0.410 \pm 0.044$ & [18] & $0.76$ & $0.440\pm 0.040$ & [19] &  $1.05$ & $0.280\pm 0.080$ & [19] \\
    $0.32$ & $0.427\pm 0.056$ & [20] &  $0.57$ & $0.426\pm 0.029$ & [20] & $0.727$ & $0.296 \pm 0.0765$ & [21]\\
    $0.02$ & $0.428\pm 0.0465$ & [22] & $0.60$ & $0.550\pm 0.120$ & [23]  & $0.86$ & $0.400\pm 0.110$ & [23] \\
    $0.1$ & $0.48 \pm 0.16$ & [24] & $0.001$ & $0.505 \pm 0.085$ &  [25]& $0.85$ & $0.45 \pm 0.11$ & [26]\\
    $0.31$ & $0.384 \pm 0.083$ &  [27]  & $0.36$ & $0.409 \pm 0.098$ &  [27]& $0.40$ & $0.461 \pm 0.086$ &  [27]  \\
    $0.44$ & $0.426 \pm 0.062$ &  [27] &  $0.48$ & $0.458 \pm 0.063$ &  [27]  & $0.52$ & $0.483 \pm 0.075$ &  [27]\\
    $0.56$ & $0.472 \pm 0.063$ &  [27] &  $0.59$ & $0.452 \pm 0.061$ &  [27] &  $0.64$ & $0.379 \pm 0.054$ &  [27] \\
    $0.1$ & $0.376\pm 0.038$ & [28]&  $1.52$ & $0.420 \pm 0.076$ &  [29]& $1.52$ & $0.396 \pm 0.079$ & [30]\\ 
    $0.978$ & $0.379 \pm 0.176$ &  [31] & $1.23$ & $0.385 \pm 0.099$ &  [31] & $1.526$ & $0.342 \pm 0.070$ &  [31] \\
    $1.944$ & $0.364 \pm 0.106$ &  [31]&  $0.60$ & $0.49 \pm 0.12$ &  [32] & $0.86$ & $0.46 \pm 0.09$ &  [32]\\
    $0.57$ & $0.501 \pm 0.051$ &[33]& $0.03$ & $0.404 \pm 0.0815$ &[34]&  $0.72$ & $0.454 \pm 0.139$ &  [35] \\
    \hline
  \end{tabular}
\end{table*}

\subsubsection{\texorpdfstring{$\hat{f}$}{fhat} measurement}\label{subsubsec:fhat}

Firstly, we provide the complete set of \(\hat{f}\) measurements utilized in our analysis in Table \ref{tab:data-rsd}, which includes 66 \(\hat{f}\) measurements from various LSS spectroscopic redshift surveys, covering the redshift range from \(z = 0.001\) to \(z = 1.944\) \citep{kazantzidis2018evolution,skara2020tension}.

Subsequently, we employed a fitting approach, Artificial Neural Network (ANN) fitting to derive the \(\hat{f}\) values for the MagLim sample at their corresponding effective redshifts. Here, we opted for ANN over the more conventional Gaussian Processes (GP) due to concerns raised in previous studies \citep{perenon2021multi}, which suggest that the current data’s distribution and precision can cause GP reconstructions to be overly sensitive to prior assumptions, such as hyperparameter ranges or mean functions. In contrast, ANN, being a machine learning method, has been demonstrated to be a universal approximator capable of modeling a wide range of functions. It is a purely data-driven technique that does not impose Gaussian assumptions. Thus, with an appropriately chosen network architecture, ANN can provide an accurate representation of the input data's distribution. Therefore, in our analysis, we used the open-source package \texttt{ReFANN}\footnote{\url{https://github.com/Guo-Jian-Wang/refann}} \citep{wang2020reconstructing} to preform the ANN fitting process.

\begin{figure}[t]
    \centering
    \includegraphics[width=0.45\textwidth]{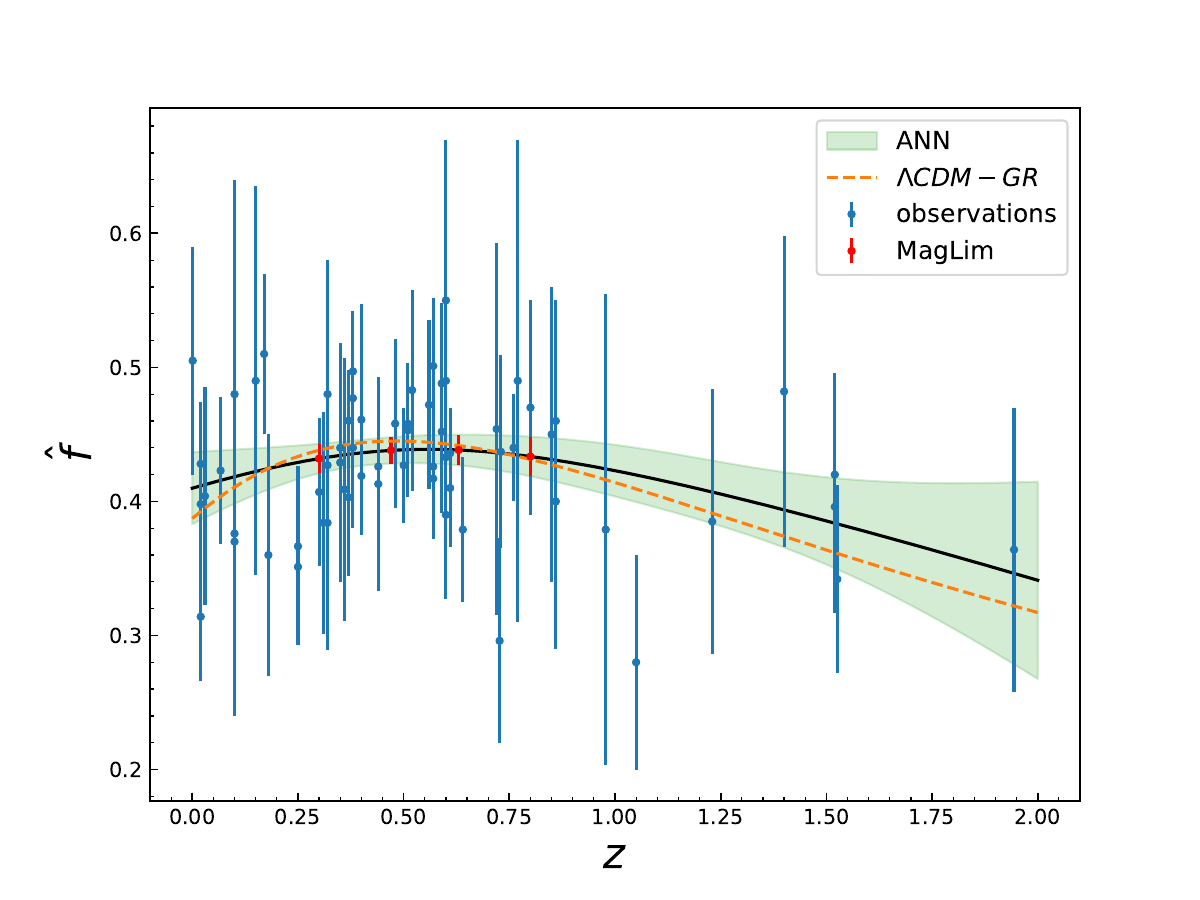}
  \caption{The reconstruction of the function \(\hat{f}(z)\) from the \(\hat{f}\) measurements, shown as the black solid line, along with the 68\% confidence level indicated by the light green region. The blue points with error bars represent the \(\hat{f}\) values obtained from the literature, while the red stars denote the inferred \(\hat{f}\) values for the DES MagLim sample. For comparison, the theoretical prediction of \(\hat{f}(z)\) based on the \(\Lambda\)CDM model is also displayed.}
  \label{fig:ann_fs8}
\end{figure}

\begin{table}[t]
  \centering
  \caption{The effective redshifts of the first four tomographic bins of DES MagLim sample, along with the corresponding galaxy bias $\hat{b} = b\sigma_8(z)$ values from \citet{abbott2023dark}, the linear growth rate $\hat{f} = f\sigma_8(z)$ values derived using \texttt{ReFANN}, and the calculated $\beta=\hat{f}/\hat{b}$ values.}
  \label{tab:values}
  \begin{tabular}{cccc}
    \toprule
    \textbf{$z_{\text{eff}}$} & \textbf{$\hat{b}= b\sigma_8(z)$} & \textbf{$\hat{f}= f\sigma_8(z)$} & \textbf{$\beta=\hat{f}/\hat{b}$} \\
    \midrule
    0.30 & $0.924\pm0.034$ & $0.432\pm0.011$ & $0.467\pm0.021$ \\
    0.47 & $0.956\pm0.044$ & $0.438\pm0.010$ & $0.458\pm0.023$ \\
    0.63 & $1.003\pm0.035$ & $0.438\pm0.011$ & $0.437\pm0.019$ \\
    0.80 & $0.865\pm0.034$ & $0.434\pm0.015$ & $0.501\pm0.026$ \\
    \bottomrule
  \end{tabular}
\end{table}
In Figure \ref{fig:ann_fs8} we present the estimations of \(\hat{f}\) for the MagLim sample at their corresponding effective redshifts. Using these 66 data points, we can determine the evolution of \(\hat{f}\) with redshift through the ANN algorithm, as shown by the black solid line in the figure. The light green region represents the 68\% confidence interval for the \(\hat{f}(z)\) provided by the ANN algorithm, and it is evident that due to the higher density of observational data between redshifts 0.3 and 0.6, the error in the final reconstruction of \(\hat{f}(z)\) is smaller around redshift 0.5. By comparing this with the theoretical curve from the \(\Lambda\)CDM model (orange dashed line), we observe that the evolution of \(\hat{f}\) derived from the existing data is consistent with that predicted by \(\Lambda\)CDM at 68\% confidence level. The final estimations of \(\hat{f}\) at four redshifts are listed in Table \ref{tab:values}.

\subsubsection{\texorpdfstring{$\hat{b}$}{bhat} measurement}\label{bhat}

We obtain the \(\hat{b}\) values for the DES MagLim sample from the data chains provided by \citet{abbott2023dark}\footnote{\url{https://desdr-server.ncsa.illinois.edu/despublic/y3a2_files/y3a2_beyond_lcdm/chains/}}. Their analysis used the first three years of observations from DES, either independently or in combination with external cosmological probes, to constrain potential extensions of the \(\Lambda\)CDM model. Specifically, they employed the two-point correlation functions of weak gravitational lensing, galaxy clustering, and their cross-correlations (commonly referred to as 3\(\times\)2pt) to constrain six different extensions of the \(\Lambda\)CDM model. 

One of these extensions is a binned \(\sigma_8(z)\) model, which serves as a phenomenological probe of structure growth without assuming specific physical mechanisms. In this approach, the binned \(\sigma_8(z)\) model is defined as:
\begin{equation}
    \sigma_8^{[\text{bin}\ i]} \equiv \sigma_8\sqrt{A_i^{P_{\text{lin}}}},
\end{equation}
where \(A_i^{P_{\text{lin}}}\) represents the amplitude of the linear matter power spectrum in the \(i\)-th redshift bin, which equals 1 in the \(\Lambda\)CDM model. Therefore, we can determine the amplitude of the growth of structure, \(\sigma_8\), in each redshift bin. Additionally, the numerical chains provide direct access to the linear bias for each redshift bin, enabling us to derive constraints on the combined parameter \(\hat{b} = b\sigma_8\). The resulting constraints on \(\hat{b}\), obtained from the chains, are presented in Table \ref{tab:values}.

In conclusion, we utilize the measurements of \(\hat{f}(z)\) and \(\hat{b}(z)\) to estimate the RSD parameter \(\beta(z)\) for the first four tomographic bins of the DES MagLim sample, with the results detailed in Table \ref{tab:values}. To assess the uncertainties in \(\beta(z)\), we apply the error propagation method. Notably, the uncertainties in our reconstructed \(\beta\) values are significantly smaller than those obtained from current constraints based on 3D power spectrum analyses. This improvement primarily stems from the reconstructed linear growth rate, which incorporates multiple \(\hat{f}\) measurements from various sources rather than relying on a single survey. Moreover, we find that the uncertainties in \(\beta\) have a minimal impact on the subsequent \(E_G\) estimates, as the dominant source of error in current \(E_G\) estimates arises from uncertainties in the power spectra.

\begin{figure*}[t]
    \centering
    \includegraphics[width=1.0\linewidth]{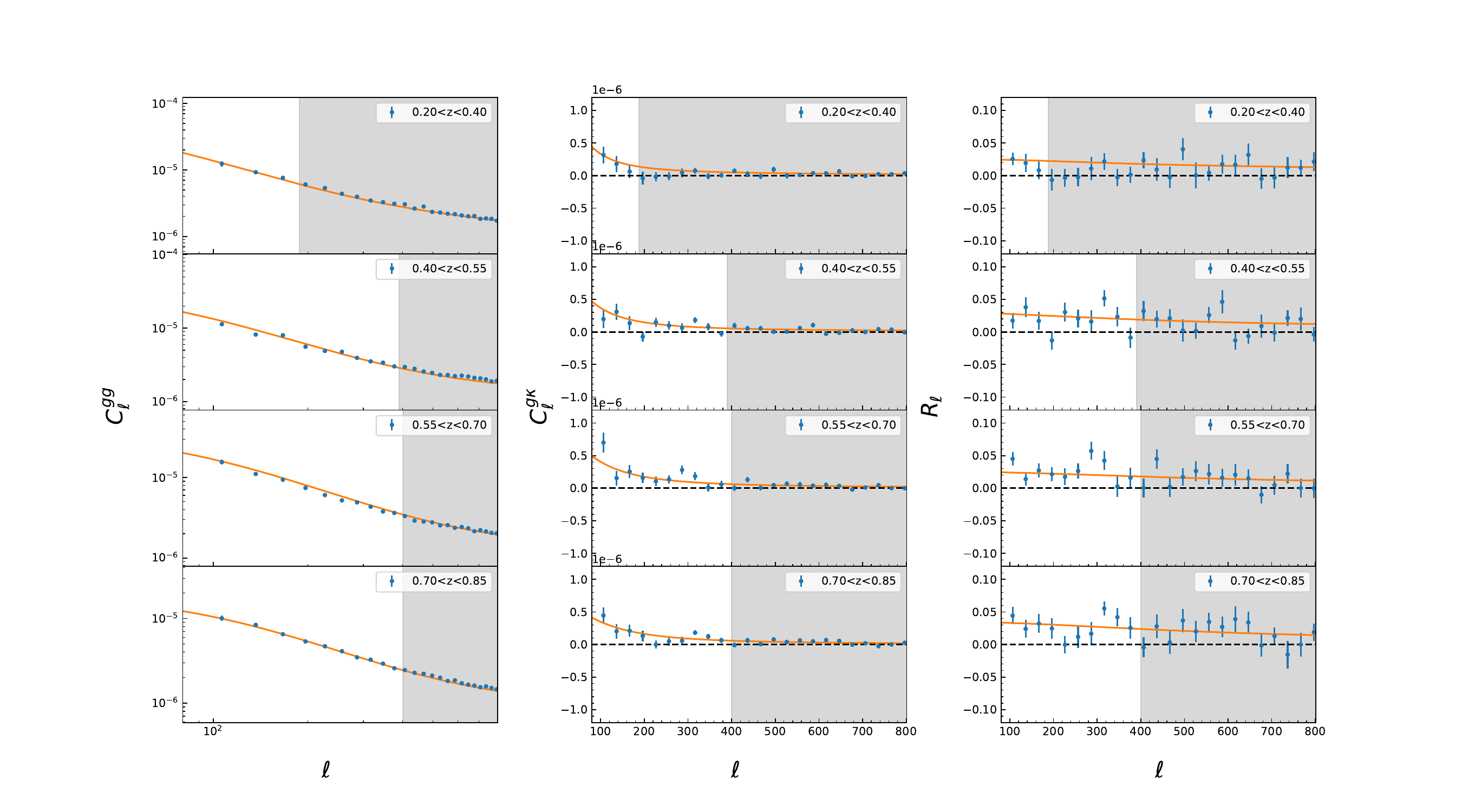}
    \caption{The observed power spectra \(C_\ell^{gg}\) (left panels), \(C_\ell^{g\kappa}\) (middle panels) and their ratio \(R_\ell\) (right panels) for the four redshift bins of the DES MagLim sample are shown. The solid orange lines depict the theoretical predictions evaluated using the best-fit parameters from \citet{abbott2023dark}. The gray-shaded regions highlight the range of multipoles that were excluded from the analysis due to the non-linearity.}
    \label{fig:cl_comp}
\end{figure*}

\subsection{Angular power spectrum}\label{subsec:cl}

We then calculate the angular power spectra \(C_\ell^{gg}\) and \(C_\ell^{g\kappa}\) using the galaxy number density map from the DES MagLim sample and the lensing measurements from Planck. These calculations are performed using the pseudo-\(C_\ell\) estimator, implemented in the \texttt{NaMaster} software package\footnote{\url{https://github.com/LSSTDESC/NaMaster}}, which provides an unbiased estimate of the angular power spectra.

The estimation of the angular power spectrum is affected by the survey's sky coverage, as the mask introduces coupling between different modes of the true power spectrum. In this framework, the true underlying power spectrum, \(C_{\ell}^{\text{true}}\), is derived from the observed power spectrum, \(C_{\ell}^{\text{obs}}\), by applying the inverse of the mode-coupling matrix, \(\text{M}\), as follows:
\begin{equation}
    C_{\ell}^{\text{true}} = \sum_{\ell^{\prime}} M_{\ell\ell^{\prime}} C_{\ell^{\prime}}^{\text{obs}}~.
\end{equation}
The mode-coupling matrix \(M_{\ell\ell^{\prime}}\) is determined entirely by the mask information. As shown in \citet{hivon2002master}, the coupling matrix \(M_{\ell\ell^{\prime}}\) can be efficiently and analytically computed due to the orthogonality of the Wigner 3j symbols. 
To mitigate the effects of the mask, we apply a binning process to the resulting power spectrum, using a wide multipole bin width of \(\Delta \ell = 30\). This approach helps to reduce the correlations between different multipole bins, ensuring that the impact of mode coupling is minimized and that the correlations among the binned multipoles remain small.

Given the sky coverage of the DES survey and the use of the Limber approximation in our analysis, we apply a conservative cut on large scales by setting the minimum multipole value to \(\ell_{\text{min}} = 80\). To address the effects of nonlinearity, we define a maximum wavenumber, \(k_{\text{nl}}\), for the transition from linear to quasi-linear scales. This is determined by the condition \(k_{\text{nl}}^{3}P(k_{\text{nl}},\bar{z})/(2\pi^{2}) = 1\), following the approach outlined in \citet{yang2018calibrating}. Given that the DES photometric data are used, we adopt a more conservative approach by setting an upper limit of \(\ell_{\text{max}} < 400\). This results in maximum multipole values of 188, 390, 400, and 400 for the respective MagLim redshift bins used in the \(E_G\) estimator. This conservative selection helps to mitigate the impact of nonlinearity and ensures the robustness of the analysis on middle scales.

Furthermore, the observed galaxy auto-correlation power spectrum is affected by the discrete nature of galaxies, which introduces an additional shot-noise contribution. Typically, the galaxy sample is assumed to follow a Poisson distribution, allowing the shot noise to be estimated as \citep{garcia2021growth}
\begin{equation}
    N_{\ell} = \sum_{\ell^{\prime}}(M^{-1})_{\ell\ell^{\prime}}\tilde{N_{\ell}},~    ~\tilde{N_{\ell}} = \frac{\langle m \rangle}{\bar{n}_{\text{eff}}}.
\label{eq:SN}
\end{equation}
Here, \(\langle m \rangle\) represents the mean value of the mask across the full sky, and \(\bar{n}_{\text{eff}}\) denotes the effective mean number density, defined as:
\begin{equation}
\bar{n}_{\text{eff}} = \frac{(\sum_{i \in p} w_i)^2}{\Omega_{\text{pix}} \sum_p m_p \sum_{i \in p} w_i^2},
\end{equation}
where \(\Omega_{\text{pix}}\) is the pixel area in steradians, \(m_p\) is the mask value for pixel \(p\), and \(w_i\) is the weight assigned to each galaxy in pixel \(p\).

In our analysis, we subtract the shot-noise contribution from the observed galaxy auto-correlation power spectrum \(C_{\ell}^{\mathrm{gg}}\). Afterward, we correct the resulting power spectrum by dividing them by the square of the pixel window function, which is obtained using the standard HEALPix tools. This process ensures that the power spectrum is properly corrected for pixelization effects. For the observed galaxy-lensing convergence power spectrum \(C_{\ell}^{\mathrm{g\kappa}}\), since the CMB lensing convergence \(\kappa\) is a continuous field unaffected by the pixel window function and the pixel window function is not applied in the \(a_{\ell m}\) to map transformation, we only need to correct for the pixelization effects by dividing by a single pixel window function.

\begin{figure*}[t]
\centering
\begin{minipage}[t]{0.4\textwidth}
\centering
\includegraphics[width=0.95\textwidth]{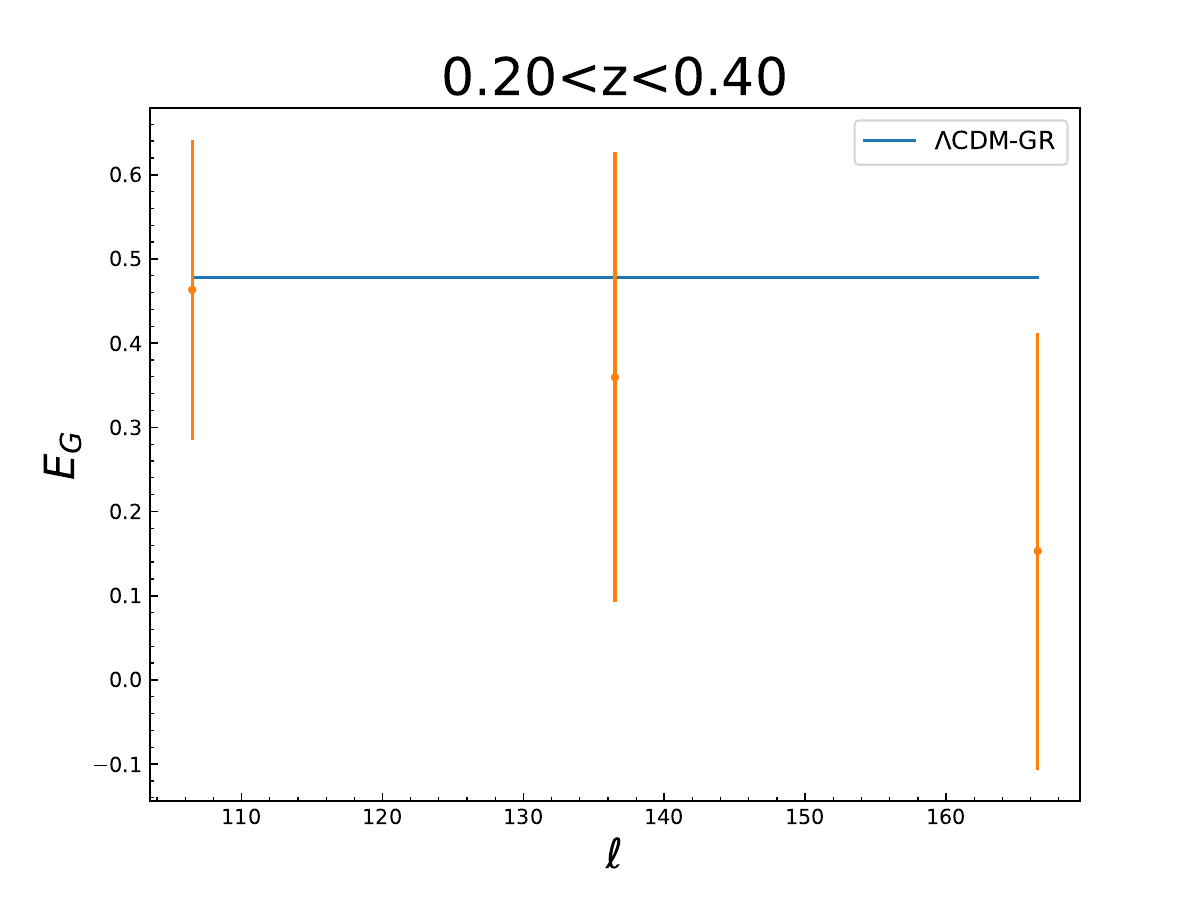}
\end{minipage}
\begin{minipage}[t]{0.4\textwidth}
\centering
\includegraphics[width=0.95\textwidth]{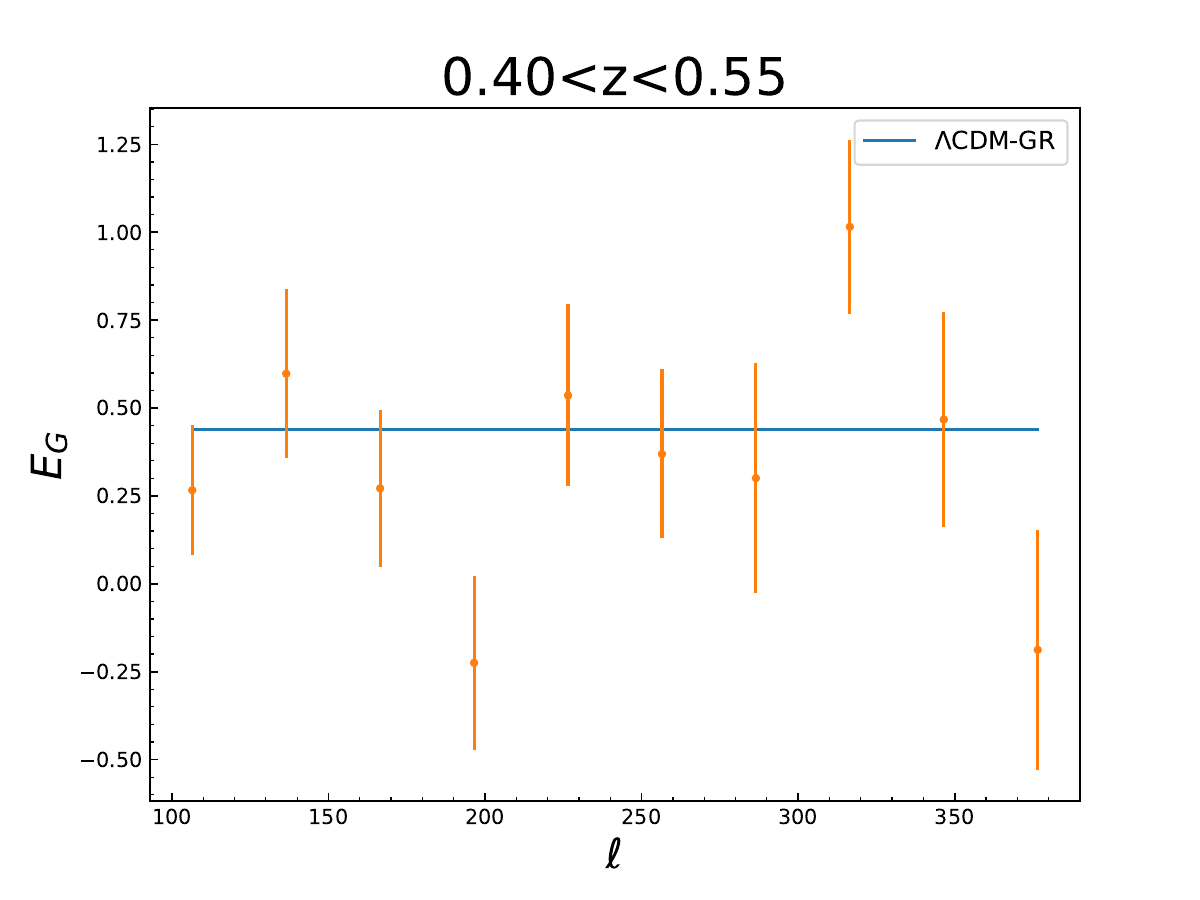}
\end{minipage}

\begin{minipage}[t]{0.4\textwidth}
\centering
\includegraphics[width=0.95\textwidth]{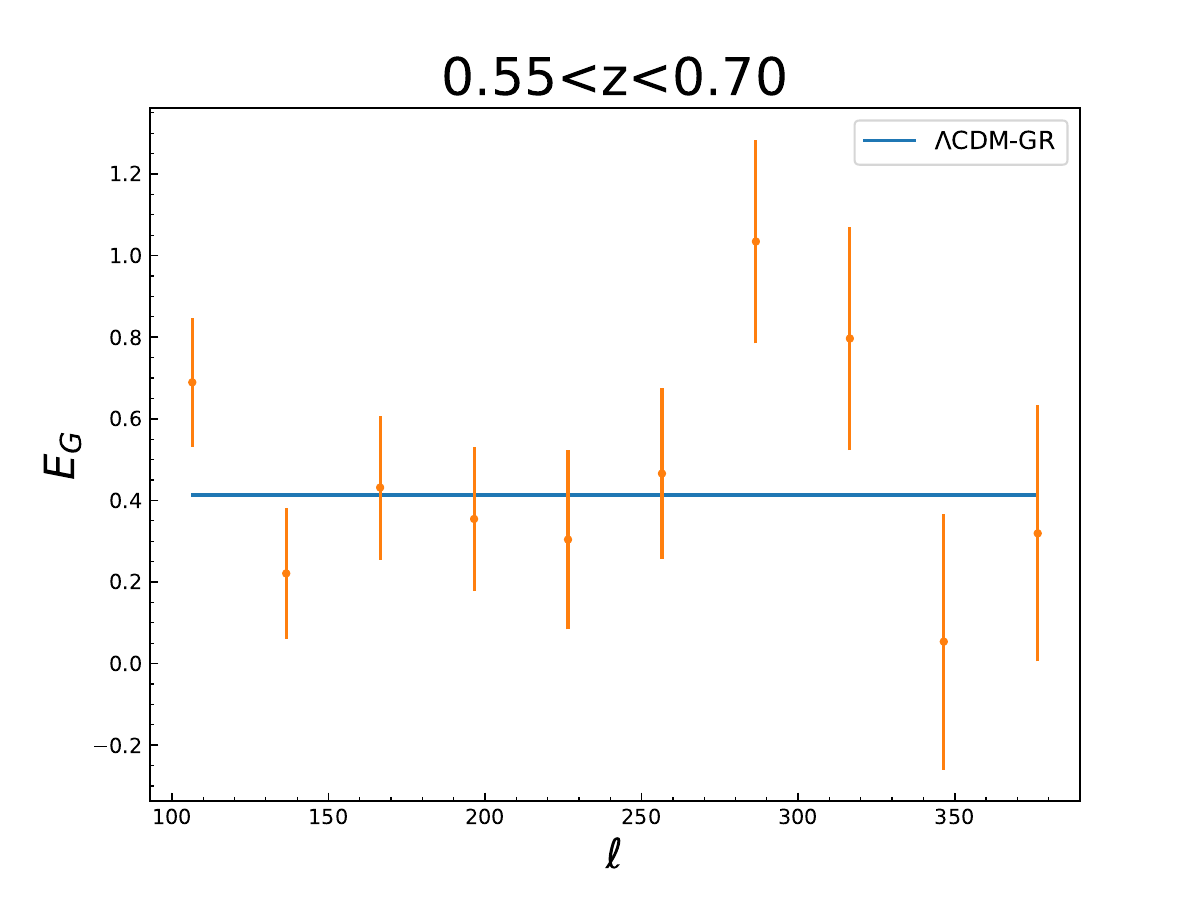}
\end{minipage}
\begin{minipage}[t]{0.4\textwidth}
\centering
\includegraphics[width=0.95\textwidth]{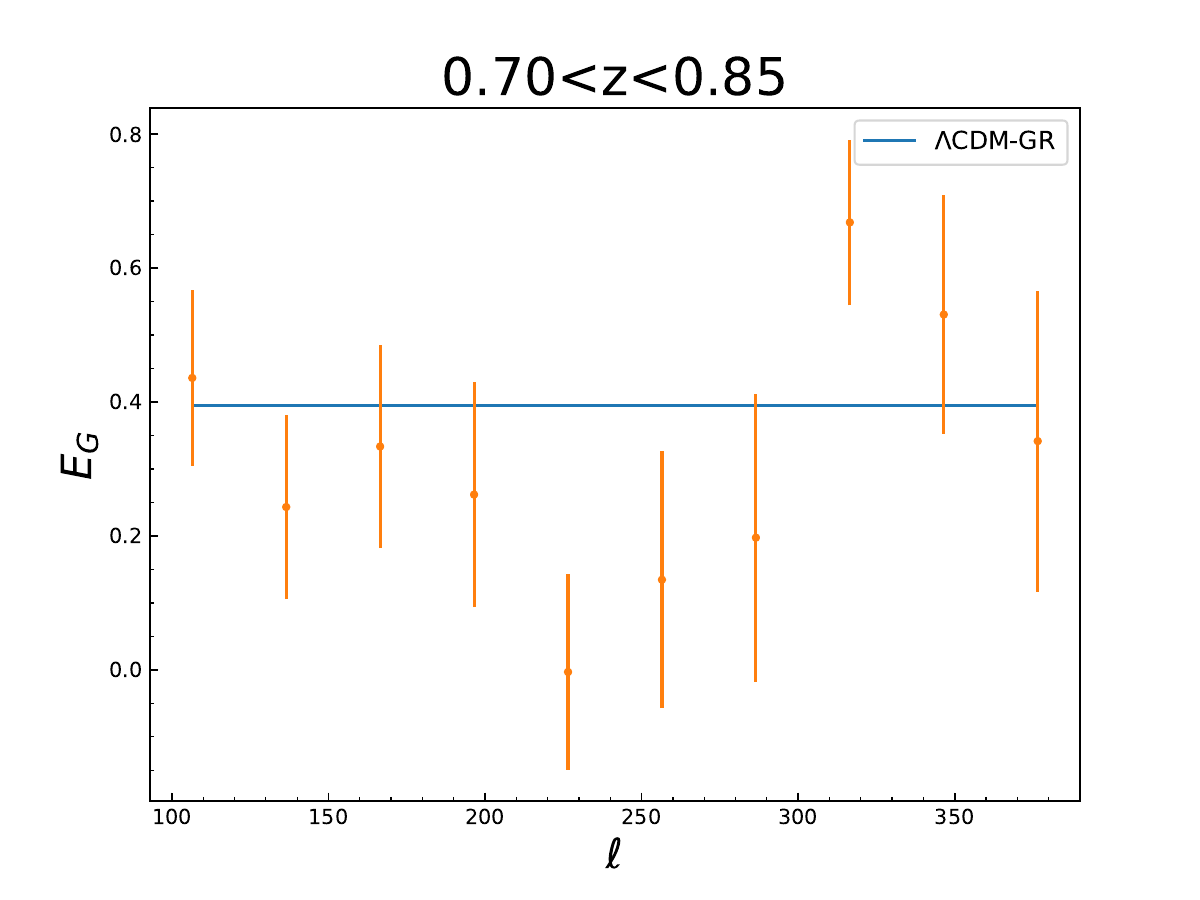}
\end{minipage}
\caption{The estimations of the \(E_G\) statistic for the DES MagLim samples, combined with the Planck lensing measurements at four redshift bins. The blue lines represent the theoretical prediction within the \(\Lambda\)CDM model, which is scale independent.}
\label{fig:maglim_cl_eg}
\end{figure*}

\subsection{Covariance matrices}\label{subsec:cov}

In our analysis, we employ the Jackknife resampling method (JK) to estimate covariance matrices, relying exclusively on the available data sample. 

The overlapping sky region between the DES MagLim sample and the Planck lensing data is partitioned into \(N_{\text{JK}}\) equal-area patches. Jackknife subsamples are generated by sequentially omitting one patch at a time, resulting in a set of leave-one-out samples. The JK-estimated mean power spectrum across all patches is calculated as follows:
\begin{equation}
\left<X_{\ell}\right> = \frac{1}{N_{\text{JK}}} \sum_{k=1}^{N_{\text{JK}}} X_{\ell}^{(k)}.
\end{equation}
The covariance of the measurements is determined by aggregating the variations across all patches. The general expression for the covariance between two angular power spectra, \(X_{\ell}\) and \(Y_{\ell^{\prime}}\), is given by:
\begin{equation}
\mathbf{C}(X_{\ell}, Y_{\ell^{\prime}}) = \frac{N_{\text{JK}} - 1}{N_{\text{JK}}} \sum_{k=1}^{N_{\text{JK}}} \left( X_{\ell}^{(k)} - \left<X_{\ell}\right> \right) \left( Y_{\ell^{\prime}}^{(k)} - \left<Y_{\ell^{\prime}}\right> \right),
\label{eq:jk}
\end{equation}
where \(X_{\ell}, Y_{\ell} \in [{C}_{\ell}^{\mathrm{g\kappa}}, {C}_{\ell}^{\mathrm{gg}}]\). 

In the case of auto-correlations, it is necessary to account for variations in shot noise within the galaxy map when a patch is removed. As outlined by \citet{wenzl2024constraining}, the shot noise in the sub-sample can be analytically represented as:
\begin{equation}
N_{\ell}^{(k)} = \frac{N_{\text{JK}}}{N_{\text{JK}} - 1} N_{\ell}.
\end{equation}

The number of jackknife samples is determined by the largest scales we aim to investigate. \citet{pullen2015probing} suggested that the minimum patch size should adequately capture the largest cosmological scales. Adhering to this recommendation, we conservatively select \(N_{\text{JK}} = 30\) patches for our analysis. We also varied the number of patches $N_{\text{JK}}$ to ensure the stability of the covariance matrix.

For a multivariate Gaussian vector with a finite sample size, the estimated covariance matrix \(\hat{\mathbf{C}}\) follows a Wishart distribution, providing an unbiased estimate of the true covariance matrix \(\mathbf{C}\). However, the inverse of the estimated covariance matrix, \(\hat{\mathbf{C}}^{-1}\), which follows an inverse Wishart distribution, is a biased estimate of the true inverse covariance matrix \(\mathbf{C}^{-1}\). This bias arises from inaccuracies in \(\hat{\mathbf{C}}\). To address this issue, we follow the approach outlined by \citet{hartlap2007your} and apply a correction factor to obtain the unbiased inverse covariance:
\begin{equation}
\hat{\mathbf{C}}_{\text{unbiased}}^{-1} = \left( 1 - \frac{N_d + 1}{N_{\text{JK}} - 1} \right) \hat{\mathbf{C}}^{-1},
\label{eq:c_unbiased}
\end{equation}
where \(N_d\) denotes the number of bandpowers utilized, and its values are 3, 10, 10, and 10 for each redshift bin.  Furthermore, \citet{dodelson2013effect} emphasized that, in maximum likelihood fitting, the errors in the inverse covariance matrix \(\hat{\mathbf{C}}^{-1}\) propagate to the model parameters. This effect can be mitigated by multiplying the inverse covariance matrix by the following factor \citep{percival2014clustering}:
\begin{equation}
\mathbf{M} = \frac{1 + B(N_d - N_p)}{1 + A + B(N_p + 1)},
\label{eq:m}
\end{equation}
where \(N_p\) represents the number of parameters, and for the estimation of the single parameter $E_G$, $N_p=1$. The constants \(A\) and \(B\) are defined as:
\begin{equation}
\begin{split}
    A &= \frac{2}{(N_{\text{JK}} - N_d - 1)(N_{\text{JK}} - N_d - 4)}, \\
    B &= \frac{N_{\text{JK}} - N_d - 2}{(N_{\text{JK}} - N_d - 1)(N_{\text{JK}} - N_d - 4)}.
\end{split}
\end{equation}

In Figure \ref{fig:cl_comp}, we present the observed power spectra \(C_\ell^{gg}\) and \(C_\ell^{g\kappa}\) derived from the DES MagLim sample and the Planck lensing measurements across four tomographic redshift bins as well as their ratio, \(R_\ell\). For illustrative purposes, we include square roots of the diagonal elements of the JK covariance matrix as \(1\sigma\) error bars on the data points. Additionally, we indicate the scale cuts, where multipoles larger than the maximum \(\ell\) are excluded, by shading the corresponding regions in gray.

Furthermore, we also conduct a comparison between the observed power spectra \(C_\ell^{gg}\) and \(C_\ell^{g\kappa}\) and their corresponding theoretical predictions for each tomographic redshift bin. The theoretical predictions are generated using the DESC Core Cosmology Library (\texttt{CCL}, \citet{chisari2019core}), with the underlying 3D power spectra computed via the CAMB Boltzmann code. To ensure consistency between the observed and theoretical power spectra, we apply the appropriate binned mode-coupling matrix and binning scheme to the theoretical curves, accounting for the effects of the survey mask and binning process. For the linear bias in each redshift bin, we adopt the best-fit values derived from the \(3\times2\)pt measurements of the DES Y3 observations \citep{abbott2023dark}, consistent with the values used in Section \ref{bhat}. The comparison shows that the measurements of both \(C_\ell^{gg}\) and \(C_\ell^{g\kappa}\) align well with the theoretical predictions.

\subsection{\texorpdfstring{$E_G(\ell)$}{eg} estimation}\label{subsec:egl}

After obtaining the measurements of the RSD parameter \(\beta\), along with the angular power spectra \(C_{\ell}^{\mathrm{gg}}\) and \(C_{\ell}^{\mathrm{g\kappa}}\), we estimate the \(E_G\) statistic as a function of the multipole using Eq.(\ref{eq:EG}), as presented in Figure \ref{fig:maglim_cl_eg}. Due to the maximum multipole for the first redshift bin being \(\ell_{\rm max} = 180\), only three multipole bins are included in the plot. To estimate the uncertainty in the \(E_G\) statistic, we employ the JK covariance matrix method, calculating the \(E_G\) values for the \(N_{\rm JK} = 30\) JK subsamples used in Section \ref{subsec:cov}. The covariance matrix of \(E_G\) is then estimated using Eq.(\ref{eq:jk}). For clarity, we plot the square roots of the diagonal elements of this covariance matrix as \(1\sigma\) error bars on the data points in Figure \ref{fig:maglim_cl_eg}.

We observe that the estimates of \(\hat{E}_G(\ell)\) are generally consistent with the predictions of GR at 68\% confidence level across most redshift bins, with no significant scale-dependent deviations. However, the \(E_G\) measurements display considerable fluctuations, suggesting that the precision of the current observational data is limited. Given that our analysis assumes the scale independence of the RSD parameter \(\beta\), these fluctuations in the \(\hat{E}_G(\ell)\) estimates are primarily influenced by the ratios of the power spectra. As shown in Figure \ref{fig:cl_comp}, even within the redshift bin \(0.55<z<0.7\), the auto-correlation power spectrum is slightly lower than the theoretical curve, the galaxy-CMB lensing cross-correlation power spectrum deviates significantly from the theoretical model for all bins in certain bandpowers. In comparison with the ratio \(R_\ell\) and \(\hat{E}_G(\ell)\) in Figure \ref{fig:maglim_cl_eg}, we find that these deviations closely correspond to the trends observed in \(R_\ell\) and \(\hat{E}_G(\ell)\) for the respective redshift bins. Specifically, the multipoles where \(R_\ell\) and \(\hat{E}_G(\ell)\) yields negative results are precisely those where \(C_{\ell}^{\mathrm{g\kappa}}\) also takes negative values. This suggests that the cross-correlation power spectrum between galaxies and CMB lensing is the dominant contributor to the observed fluctuations. To validate this idea, we roughly estimated the contributions to the \( E_G \) error from the three components (\(C_{\ell}^{\mathrm{gg}}, C_{\ell}^{\mathrm{g\kappa}}, and\ \beta\)) using the error propagation formula. For simplicity, we assumed the three components to be independent, and the results indicated that the cross-correlation contributes over 90\% to the \( E_G \) error. Therefore, improving the precision of \(\hat{E}_G(\ell)\) estimates would likely benefit significantly from more accurate measurements of the cross-correlation power spectrum \(C_{\ell}^{\mathrm{g\kappa}}\).

In contrast to the first three redshift bins, it is notable that the \(E_G\) results within the redshift range \(0.7 < z < 0.85\) are generally lower than those predicted by the \(\Lambda\)CDM model, a trend that becomes more pronounced in the subsequent constant \(E_G\) estimation. However, as shown in Figure \ref{fig:cl_comp}, the power spectra \(C_{\ell}^{\mathrm{gg}}\) and \(C_{\ell}^{\mathrm{g\kappa}}\) for this redshift bin align closely with theoretical predictions, suggesting that this discrepancy may be due to an overestimated \(\beta\) parameter. We posit that the primary cause of the overestimation of \(\beta(z)\) stems from the underestimation of \(\hat{b}(z)\), as derived from the DES chains in the fourth redshift bin. Specifically, the best-fit value of \(\hat{b}(z_4)\) is noticeably lower compared to the values obtained in the other three redshift bins. Although, in theory, \(E_G\) should be independent of galaxy bias, as this term cancels out in the calculation, this assumption holds only if \(C_{\ell}^{\mathrm{gg}}\), \(C_{\ell}^{\mathrm{g\kappa}}\), and \(\beta\) yield consistent estimates of galaxy bias. However, these three constraints may not be fully consistent.

For instance, \citet{marques2024cosmological} analyzed the constraints on galaxy bias from the auto-correlation of MagLim galaxies and their cross-correlation with CMB lensing using the fourth data release of the Atacama Cosmology Telescope (ACT). Their findings indicate that, with other cosmological parameters fixed, \(C_{\ell}^{\mathrm{gg}}\) favors a higher galaxy bias than \(C_{\ell}^{\mathrm{g\kappa}}\), with the difference reaching 2.43\(\sigma\) in the \(0.7 < z < 0.85\) bin. This discrepancy likely impacts the constraint on \(\hat{b}\), derived from the 3\(\times\)2pt analysis, which may also be subject to such inconsistencies. Therefore, we suggest that this inconsistency could be a significant factor affecting the final \(E_G\) estimates in this redshift range.

\subsection{Scale Independent \texorpdfstring{$E_G$}{eg}}\label{subsec:egz}

In addition to examining the \(E_G\) statistic as a function of multipoles, we also consider the estimation of a scale-independent \( \bar{E}_G \) by fitting a constant value across all scales. The best-fit value of \( \bar{E}_G \) is inferred by minimizing the \( \chi^2 \) function, given by 
\begin{equation}
    \chi^2 = \left[ \hat{E}_G(\ell) - \bar{E}_G \right]^{T} \hat{\mathbf{C}}^{-1} \left[ \hat{E}_G(\ell) - \bar{E}_G \right],
\end{equation}
where \( \hat{\mathbf{C}} \) denotes the estimated covariance matrix of \( \hat{E}_G(\ell) \), which is also estimated using the jackknife resampling method. The maximum likelihood estimate for \( \bar{E}_G \) can be expressed analytically, as shown in \citet{zhang2021testing}:
\begin{equation}
    \bar{E}_G = \frac{\sum_{\ell,\ell^{\prime}} \hat{\mathbf{C}}_{\ell\ell^{\prime}}^{-1} \hat{E}_G(\ell^{\prime})}{\sum_{\ell,\ell^{\prime}} \hat{\mathbf{C}}_{\ell\ell^{\prime}}^{-1}},
\end{equation}
with the corresponding statistical uncertainty
\begin{equation}
    \sigma(\bar{E}_G) = \mathbf{M} \times \left( \sum_{\ell,\ell^{\prime}} \hat{\mathbf{C}}_{\ell\ell^{\prime}}^{-1} \right)^{-1/2},
\end{equation}
where \( \hat{\mathbf{C}}_{\ell\ell^{\prime}}^{-1} \) represents the unbiased covariance matrix obtained in the previous section, and \( \mathbf{M} \) is the correction factor introduced in Eq.(\ref{eq:m}).

\begin{figure}[t]
    \centering
    \includegraphics[width=1.0\linewidth]{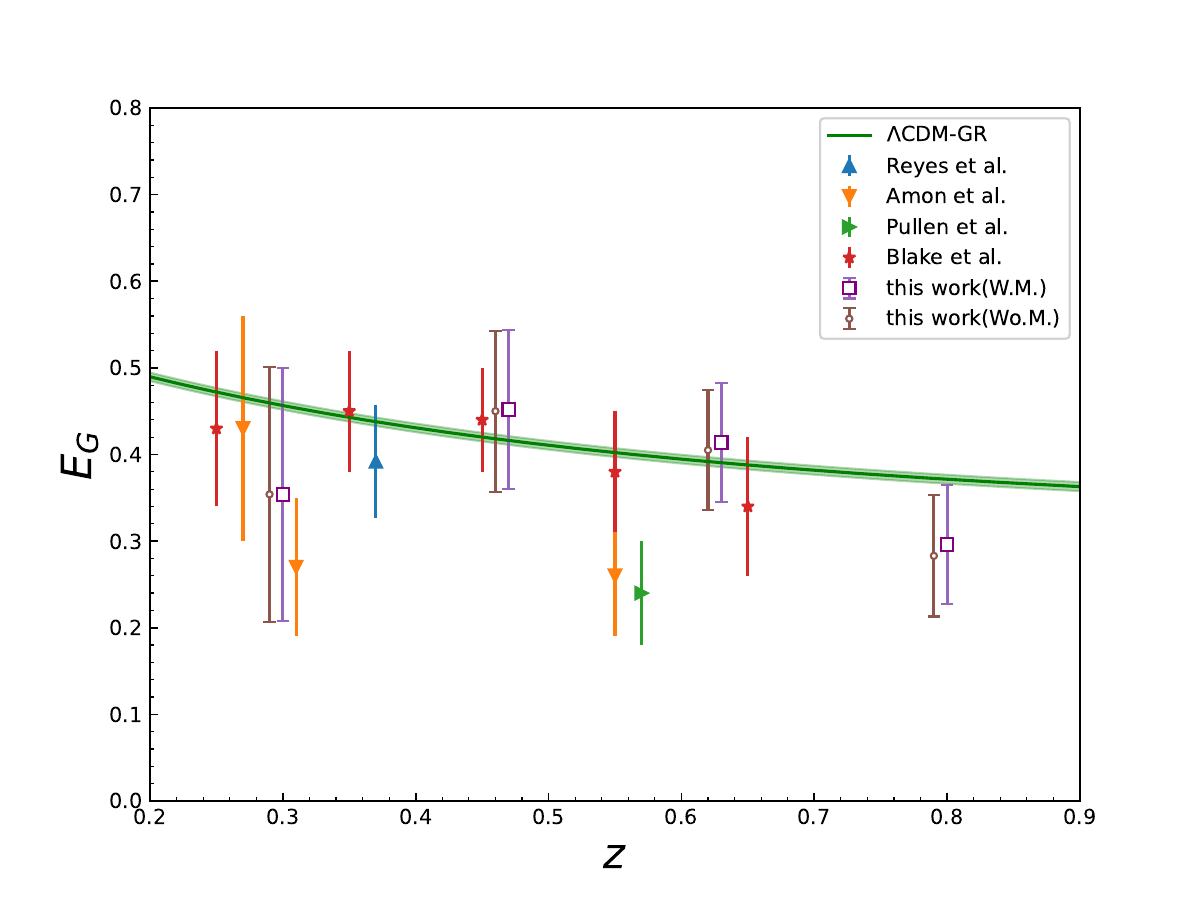}
    \caption{The scale-independent measurements of the \(E_G\) statistic for the four redshift bins of the DES MagLim sample. W.M. and Wo.M. represent measurements with and without the magnification bias effect, respectively. For comparison, previous \(E_G\) measurements from other studies are also shown, alongside the theoretical prediction based on the \(\Lambda\)CDM model (green solid line).}
    \label{fig:eg_z_cmb}
\end{figure}

Finally, we obtain the measurements of scale independent $E_G$ in four redshift bins at 68\% confidence level:
\begin{equation}
\begin{split}
    \bar{E}_G(z_1) & = 0.354\pm0.146 \\
    \bar{E}_G(z_2) & = 0.452\pm0.092 \\
    \bar{E}_G(z_3) & = 0.414\pm0.069 \\
    \bar{E}_G(z_4) & = 0.296\pm0.069    
\end{split}
\end{equation}
In Figure \ref{fig:eg_z_cmb}, we present a summary of previous \(E_G\) estimates alongside the results obtained in this work, compared with the predicted values of \(E_G\) within the \(\Lambda\)CDM model. Our measurements show good consistency with other \(E_G\) estimates, as well as with the \(\Lambda\)CDM predictions. However, our results exhibit larger statistical uncertainties, particularly at lower redshifts. This is likely due to the smaller \(\ell_{\text{max}}\) chosen in these redshift bins, which reduces the amount of available information.
Comparing different bins indicates that larger \(\ell_{max}\) values generally lead to smaller statistical uncertainties in \(E_G\). However, even with increased \(\ell_{max}\), the errors remain too large to effectively differentiate between various modified gravity models.

Finally, we conducted a brief analysis of the impact of magnification bias on our results. \citet{dizgah2016lensing} highlighted that foreground density perturbations not only magnify the regions around galaxies but also alter the selection of galaxies near the observational flux limit, potentially introducing biases in galaxy clustering. In subsequent studies, \citet{yang2018calibrating} found that the influence of magnification bias is relatively minor for most spectroscopic galaxy surveys but becomes significant in photometric galaxy surveys. To qualitatively assess this impact, we adopted the magnification bias parameters from \citet{marques2024cosmological}, applying values of 0.642, 0.63, 0.776, and 0.794 for the four redshift bins, respectively.

We then computed the theoretical changes in \(C_{\ell}^{\mathrm{gg}}\), \(C_{\ell}^{\mathrm{g\kappa}}\), and \(E_G\) with and without accounting for this magnification bias correction. Our findings suggest that the effect of magnification bias on the auto-correlation power spectrum is negligible, while its impact on the cross-correlation power spectrum increases with redshift but remains modest. The influence on \(E_G\) is most pronounced in the highest redshift bin, resulting in an average shift of approximately 4\%. The results are also presented in Figure \ref{fig:eg_z_cmb}, with labels Wo.M. Given the current level of uncertainty in our measurements, we conclude that the effect of magnification bias can be considered negligible for this analysis.

\section{forecast}\label{sec:forecast}

Given that the current $E_G$ measurements derived from existing observations lack the precision required to effectively distinguish between different MG models, we now turn to an exploration of potential improvements in the estimation of $E_G$ with upcoming LSS photometric surveys and CMB measurements. Future surveys, such as the China Space Station Telescope (CSST), promise to significantly enhance both the quantity and quality of photometric data, offering improved galaxy clustering measurements. Similarly, the CMB-S4 will yield more precise CMB lensing maps, which will contribute to reducing uncertainties in the cross-correlation power spectrum.

\subsection{CSST}\label{subsec:csst}

The CSST, as part of the Chinese Space Station Optical Survey(CSS-OS), is equipped with a two-meter aperture and seven photometric filters spanning a broad wavelength range of \(255\ \text{nm}\) to \(1000\ \text{nm}\). These filters—\(NUV\), \(u\), \(g\), \(r\), \(i\), \(z\), and \(y\)—are designed to detect point sources with 5\(\sigma\) magnitude limits ranging from 24.4 to 26.3 AB magnitudes, depending on the band. 

To characterize the number density distribution of galaxies in the survey, we adopt a parametric model for the galaxy distribution, \(n(z)\), expressed as:
\begin{equation}
n(z) \propto z^{\alpha} \exp\left[-\left(\frac{z}{z_0}\right)^{\beta}\right],
\end{equation}
where the parameters \(\alpha\), \(\beta\), and \(z_0\) are chosen to describe the survey's galaxy population. Based on previous works like \citet{gong2019cosmology}, we use \(\alpha=2\), \(\beta=1\), and \(z_0=0.3\), which roughly match the expected distribution of galaxies for photometric surveys like the CSST.

For tomographic analysis, where galaxies are divided into several redshift bins, the number density of galaxies in a specific redshift bin \(i\), denoted as \(n_i(z)\), is derived by integrating the galaxy distribution \(n(z)\) over the redshift bin limits and incorporating uncertainties from photometric redshift errors. The probability distribution of the observed redshift \(z_p\) given the true redshift \(z\) is modeled by a Gaussian distribution:
\begin{equation}
p(z_p \mid z) = \frac{1}{\sqrt{2\pi} \sigma_z} \exp\left[-\frac{(z - z_p)^2}{2 \sigma_z^2}\right],
\end{equation}
where \(\sigma_z\) is the redshift scatter, typically modeled as a function of redshift. For simplicity, we assume a constant redshift scatter \(\sigma_z = 0.05\), which is a reasonable approximation for future photometric surveys.

The effective galaxy distribution in a given redshift bin is then calculated using:
\begin{equation}
n_i(z) = \frac{1}{2} n(z) \left[\texttt{erf}\left(\frac{z_{\max} - z}{\sqrt{2} \sigma_z}\right) - \texttt{erf}\left(\frac{z_{\min} - z}{\sqrt{2} \sigma_z}\right)\right],
\end{equation}
where \(\texttt{erf}\) is the error function, and \(z_{\min}\) and \(z_{\max}\) represent the edges of the redshift bin. This equation accounts for the smearing effect of redshift errors, allowing us to compute the number density of galaxies within each tomographic bin.

For this analysis, we divide the survey into four uniform redshift bins, and the corresponding surface galaxy densities \(\bar{n}_i\) in units of galaxies per square arcminute are assigned as 7.9, 11.5, 4.6, and 3.7 for the four bins, respectively. The galaxy bias, which affects the clustering properties of galaxies, is assumed to follow a linear relation with redshift:$b(z) = 1 + 0.84z$, as derived from previous forecasts for similar surveys \citep{gong2019cosmology,lin2022forecast}. This parameterization of galaxy bias and number density will allow for accurate predictions of galaxy clustering and cross-correlation power spectra, enabling the study of cosmological parameters such as dark energy, gravity, and the growth of large-scale structure with the CSST.

\begin{figure*}[t]
\centering
\begin{minipage}[t]{0.4\textwidth}
\centering
\includegraphics[width=0.95\textwidth]{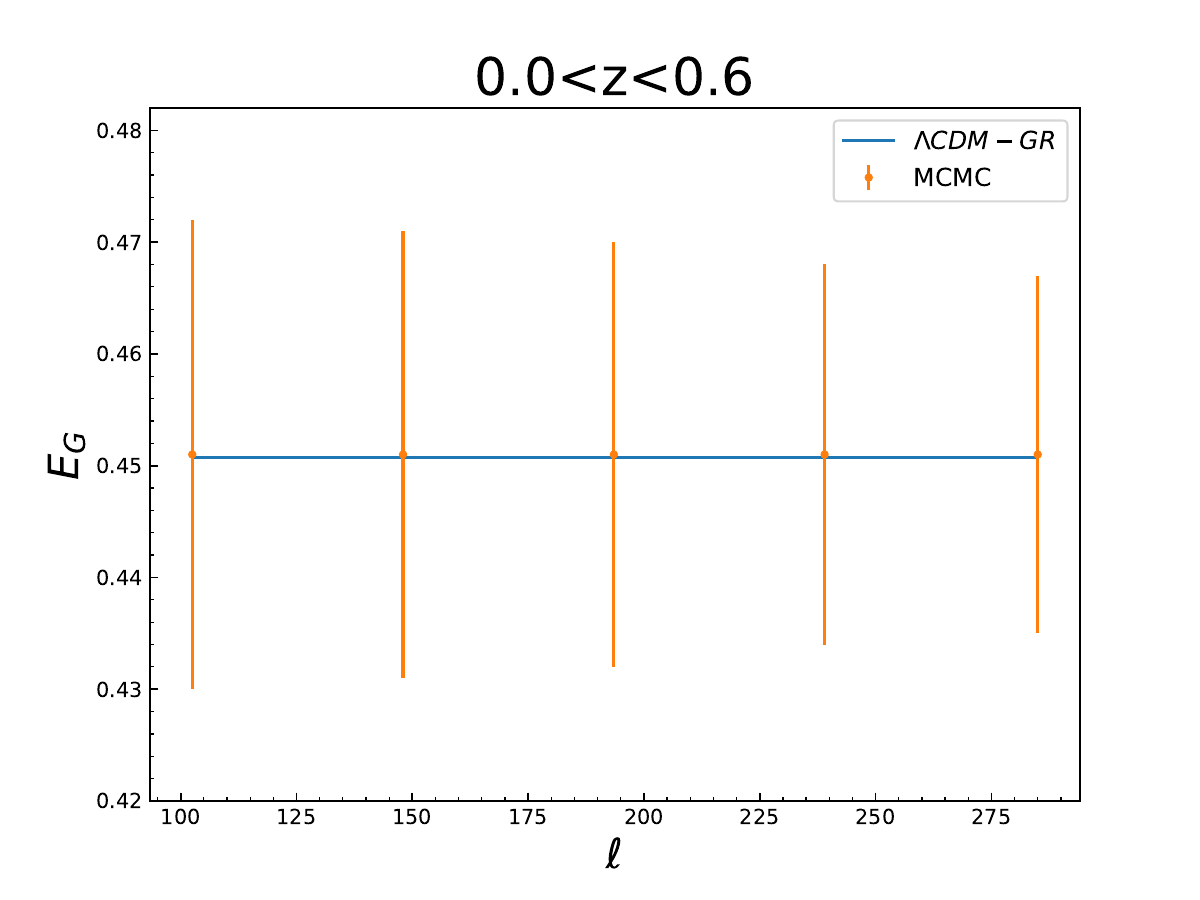}
\end{minipage}
\begin{minipage}[t]{0.4\textwidth}
\centering
\includegraphics[width=0.95\textwidth]{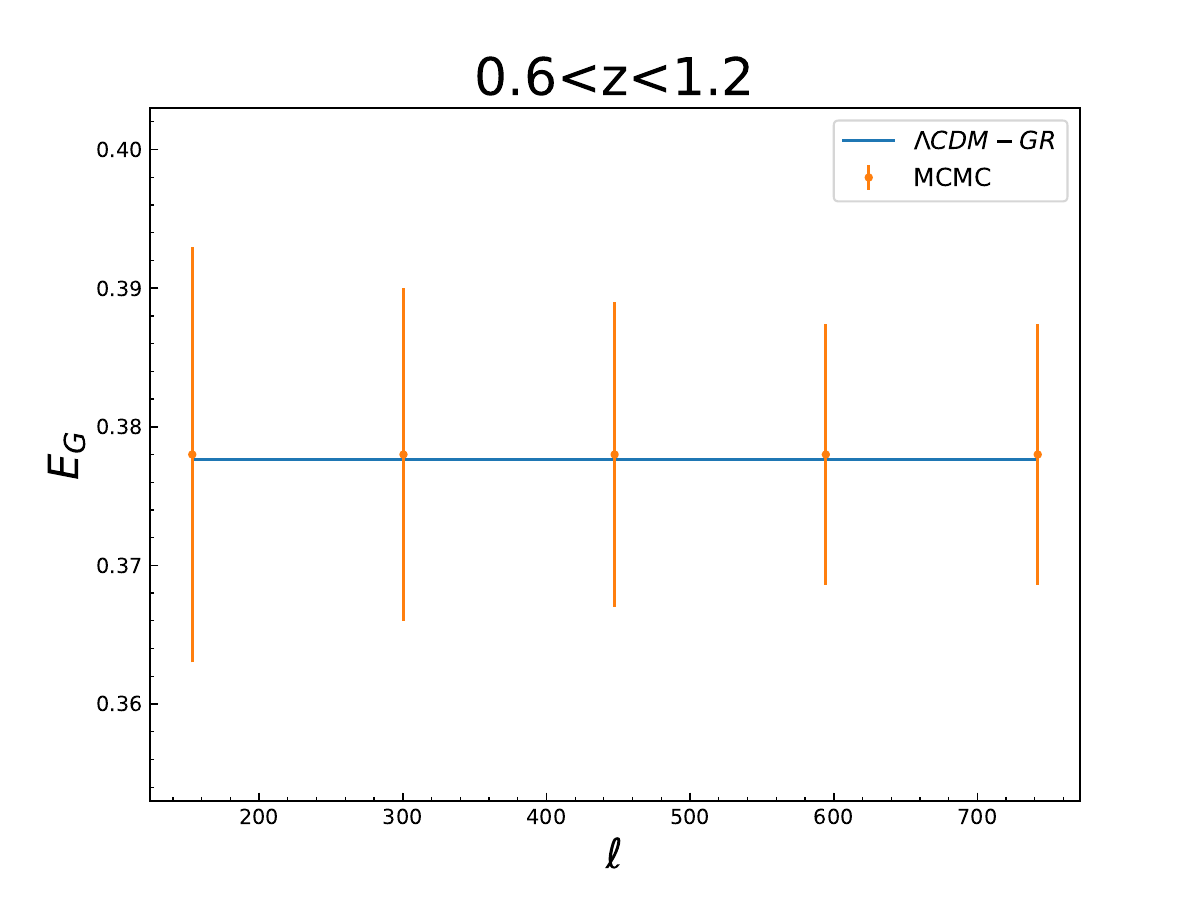}
\end{minipage}

\begin{minipage}[t]{0.4\textwidth}
\centering
\includegraphics[width=0.95\textwidth]{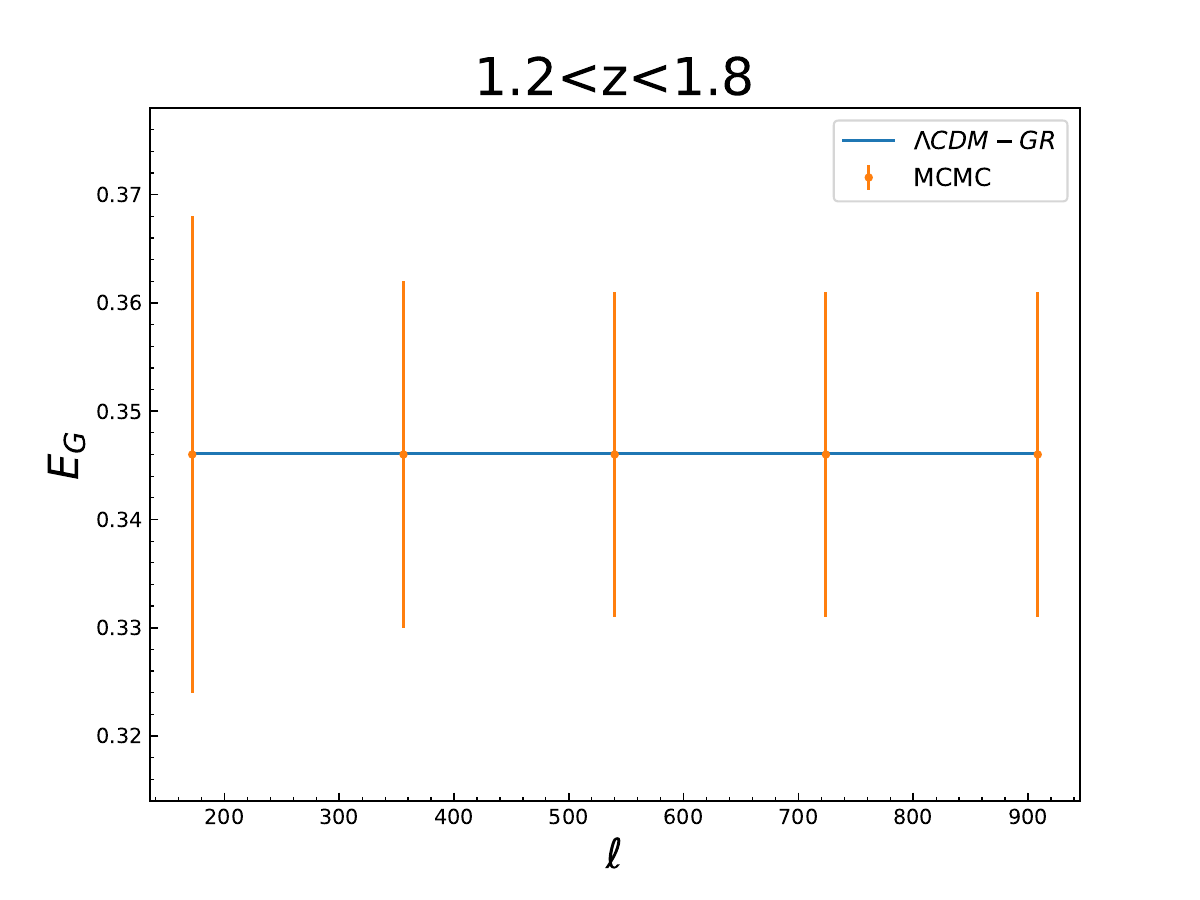}
\end{minipage}
\begin{minipage}[t]{0.4\textwidth}
\centering
\includegraphics[width=0.95\textwidth]{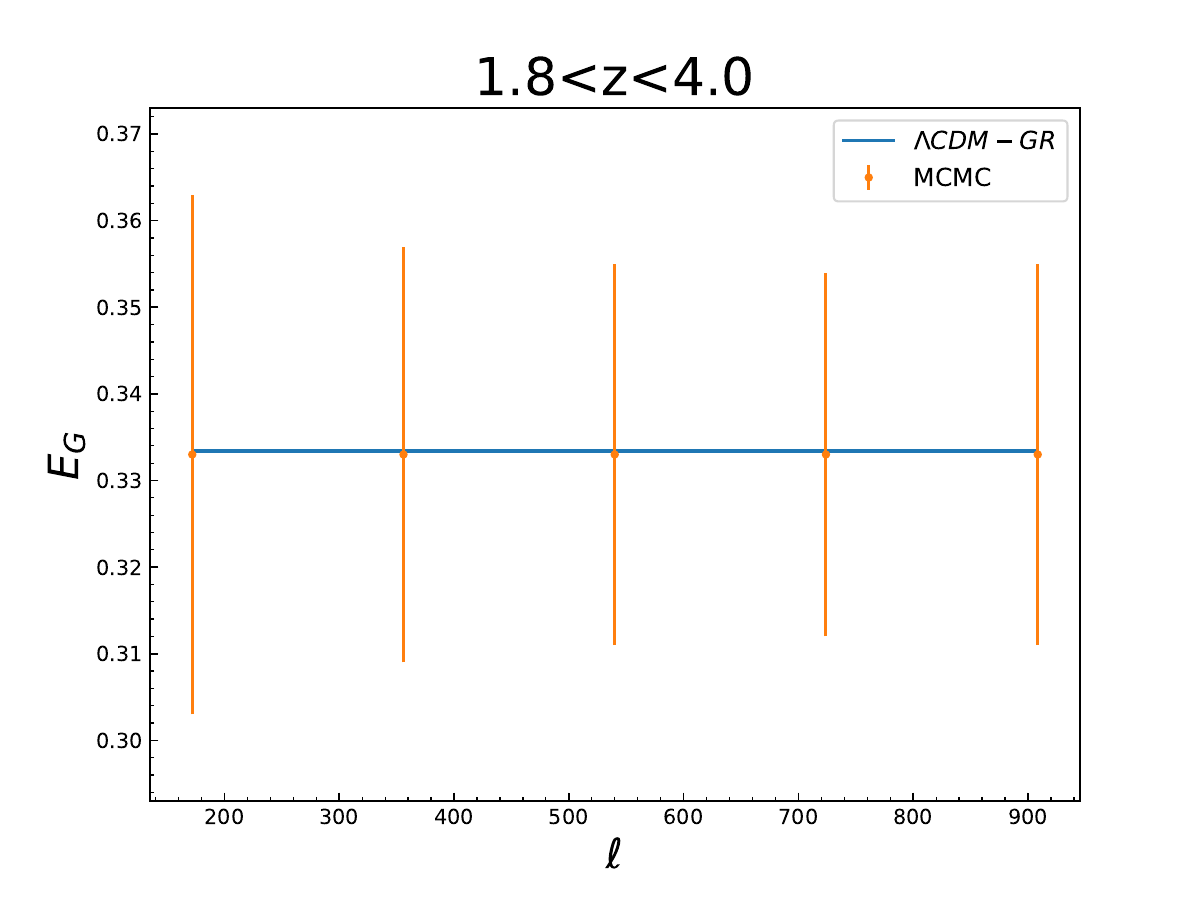}
\end{minipage}
\caption{The estimations of the \(E_G\) statistic at redshift bins for the CSST photometric redshift survey, combined with the CMB-S4 lensing measurements. The blue lines represent the theoretical prediction within the \(\Lambda\)CDM model.}
\label{fig:csst_cl_eg}
\end{figure*}

\subsection{CMB-S4}\label{subdec:s4}

The next-generation Stage-4 ground-based cosmic microwave background (CMB) experiment, CMB-S4 \citep{abazajian2016cmb,abazajian2022snowmass}, will be a major advancement in CMB science, equipped with dedicated telescopes featuring highly sensitive superconducting cameras. CMB-S4 is poised to push the boundaries of CMB lensing research by producing lensing maps with significantly higher signal-to-noise ratios, thanks to its superior sensitivity and enhanced polarization capabilities. With better polarization sensitivity, CMB-S4 will generate lensing maps less affected by contamination from foregrounds, such as dust or other cosmic sources, that can obscure CMB signals. Furthermore, the experiment's multi-frequency approach will aid in reducing foreground contamination in temperature- and polarization-based lensing estimates. Specifically, the polarization-based estimates, relying on E- and B-modes, will be the most important for CMB-S4, providing more accurate and sharper lensing maps. This will also facilitate improved cross-correlation with large-scale structure (LSS) maps from next-generation galaxy surveys.

For the purposes of analysis, we model the CMB-S4 telescope beam with a Full-Width-Half-Maximum (FWHM) of \(1'\) and assume white noise levels of \(1~\mu{\rm K}\ \text{arcmin}\) for temperature and \(1.4~\mu{\rm K} \ \text{arcmin}\) for polarization. The noise power spectra deconvolved with a beam, \(N_\ell^{\rm TT}\) for temperature and \(N_\ell^{\rm EE}\ and\ N_\ell^{\rm BB}\) for polarization, are modeled as Gaussian noise, given by the expression:
\begin{equation}
N_\ell^{\rm XX} = s_{X}^2 \exp\left[\ell(\ell+1)\frac{\theta^2_{\rm FWHM}}{8\,{\rm log}\,2}\right],
\end{equation}
where \(\text{X}\) refers to either temperature (T) or polarization (E,B), \(s_{X}\) is the polarization and temperature noises in \(\mu {\rm K}\,{\rm rad}\), and \(\theta_{\rm FWHM}\) is the beam's FWHM in radians.

For CMB lensing reconstruction, we utilize the quadratic estimator method for the EB mode (E-mode polarization and B-mode lensing signal), as described by \citet{hu2002mass}. This approach is implemented using the \texttt{QUICKLENS} software package, which allows for efficient estimation of lensing potentials. The combination of high sensitivity, low-noise levels, and advanced reconstruction techniques in CMB-S4 will enable a dramatic improvement in lensing signal extraction, offering new insights into the underlying structure and evolution of the universe.

\subsection{\texorpdfstring{$E_G$}{eg} Uncertainties}\label{subsec:result}

Based on the performance parameters of the CSST photometric redshift survey and the CMB-S4 lensing measurements, we estimate the uncertainties associated with the \(E_G\) statistic across four distinct redshift bins. Consistent with previous analyses, we establish a minimum multipole of \(\ell_{\text{min}} = 80\) and utilize the \(k_{\rm nl}\) relation to determine the maximum multipoles for each of the redshift bins. Given the anticipated performance enhancements offered by the CSST, we set the cutoff scale for \(\ell_{\text{max}}\) at 1000. Consequently, we derive the maximum multipoles for the four redshift bins as follows: \(\ell_{\rm max} = 307,\,815,\,1000,\,1000\).

In our analysis, we adopt a Markov Chain Monte Carlo (MCMC) approach to enhance our estimates. The corresponding \(\chi^2\) statistic is formulated as follows:
\begin{equation}
    \chi^2 = \sum_{\ell} \left[ \langle\vec{d}(\ell)\rangle - \vec{t}(\ell) \right] \text{Cov}^{-1} \left[ \langle\vec{d}(\ell)\rangle - \vec{t}(\ell) \right],
\end{equation}
where \(\langle\vec{d}(\ell)\rangle\) and \(\vec{t}(\ell)\) represent the averaged data vectors and theoretical vectors, respectively. In this work, the averaged data vectors (\(\langle\hat{C}_{\ell}^{\mathrm{gg}}\rangle, \langle\hat{C}_{\ell}^{\mathrm{g\kappa}}\rangle, \langle\hat{\beta}\rangle\)) include the instrument noise, while the theoretical vectors (\(C_{\ell}^{\mathrm{gg}}, C_{\ell}^{\mathrm{g\kappa}}, \beta\)) are derived from the constrained parameters summarized in Table \ref{tab:mcmc}. The total \(\chi^2\) for our analysis is estimated as:
\begin{equation}
    \chi^2_{\text{tot}} = \chi^2_{\mathrm{gg}} + \chi^2_{\mathrm{g\kappa}} + \chi^2_{\beta},
\end{equation}
where \(\chi^2_{\mathrm{gg}}\), \(\chi^2_{\mathrm{g\kappa}}\), and \(\chi^2_{\beta}\) correspond to the contributions from photometric galaxy clustering, galaxy-CMB lensing, and the RSD parameter \(\beta\), respectively.

Initially, we estimate the power spectra \(C_{\ell}^{\mathrm{gg}}\) and \(C_{\ell}^{\mathrm{g\kappa}}\) derived from the CSST photometric redshift survey and CMB-S4 lensing measurements, alongside the measurements of \(\beta\) from the CSST photometric redshift survey. Subsequently, we jointly constrain related cosmological parameters, including the current matter energy density \(\Omega_{\rm m,0}\) and the amplitude of structure growth \(\sigma_8\), utilizing these measurements. Finally, we leverage the constraints on these cosmological parameters to infer the uncertainties associated with \(E_G\) at the relevant scales.
\begin{center}
\begin{table}[t]
  \caption{Free parameters considered in the constraint process. The first column shows the names of our free parameters. The second and third columns show the fiducial values and the prior ranges of the parameters.}
  \label{tab:mcmc}
  \begin{tabular}{ccc}
    \toprule
    \textbf{Parameter} & \textbf{Fiducial Values} & \textbf{Prior}\\
    \midrule
    $\Omega_{\rm m,0}$ & 0.32 & flat(0,1)  \\
    $n_s$ & 0.9665 & flat(0.7,1)  \\
    $\sigma_8$ & 0.83 & flat(0,1)  \\
    $h_0$ & 0.72 & flat(0.5,1)  \\
    \bottomrule
  \end{tabular}
\end{table}
\end{center}

The covariance of the angular power spectra \(C_{\ell}^{\mathrm{gg}}\) and \(C_{\ell}^{\mathrm{g\kappa}}\) is expressed as follows:
\begin{equation}
\begin{aligned}
&\operatorname{Cov}\left[\tilde{C}_{\mathrm{XY}}(\ell), \tilde{C}_{\mathrm{XY}}\left(\ell^{\prime}\right)\right] \\
&=\frac{\delta_{\ell \ell^{\prime}}}{f_{\text{sky}} \Delta\ell (2 \ell + 1)} \left[\tilde{C}_{\mathrm{XX}}(\ell) \tilde{C}_{\mathrm{YY}}(\ell) + \tilde{C}_{\mathrm{XY}}(\ell) \tilde{C}_{\mathrm{XY}}(\ell)\right]~,
\end{aligned}
\label{eq:fisher_cov}
\end{equation}
where \(f_{\text{sky}} = 0.323\) represents the fraction of the sky covered, and \(\tilde{C}_{\mathrm{XY}}(\ell)\) denotes the signal of the angular power spectra augmented by shot noise. Here, \(X\) and \(Y\) may refer to the same or different tracers, such as galaxies and CMB lensing. To estimate the error in the RSD parameter \(\beta\) for each redshift bin, we employ the relation \(\sigma(\beta)/\beta = 0.085 \sqrt{0.1(1+1)/(z_2 - z_1)}\) \citep{yang2018calibrating}, where \(z_2\) and \(z_1\) are the upper and lower limits of the redshift distribution.

We employ the MCMC sampler from the publicly available \texttt{Cobaya} package \citep{torrado2021cobaya} to perform likelihood sampling. The convergence of the chains is assessed using the generalized version of the R-1 Gelman-Rubin statistic \citep{lewis2013efficient,gelman1992inference}, with convergence defined by the criterion \(R - 1 < 0.01\). To mitigate the effects of initial conditions, the first 30\% of the chains are discarded as burn-in. Upon obtaining constraints on the cosmological parameters, we utilize these constraints to infer the uncertainties associated with \(E_G\) at the relevant scales.

Figure \ref{fig:csst_cl_eg} illustrates the estimated results of \(E_G\) as a function of multipoles for various redshift bins, demonstrating consistency with the theoretical model within the \(1\sigma\) confidence interval. Notably, our estimates exhibit more than a fivefold improvement in precision compared to the photometric results obtained from the DES. Furthermore, in our estimation of the constant \(\bar{E}_G\), the overall error achieved is generally at the \(1\%\) level, representing a significant enhancement over current results. According to \citet{pullen2015probing}, this level of precision enables the differentiation of GR from chameleon gravity (with \(\beta > 1.1\)) at the \(5\sigma\) level, and allows for the distinction of \(f(R)\) gravity from GR at the \(13\sigma\) level for \(B_0 > 10^{-7}\), thereby providing a stringent test of the viability of \(f(R)\) theories.

Finally, we assessed the impact of magnification bias on our results. To simplify our analysis, we employed the magnification bias fitting formula derived from the Flagship simulations of the Euclid mission \citep{lepori2022euclid}, which has performance characteristics similar to those of the CSST. The fitting formula is expressed as follows:
\begin{equation}
s(z) = s_0 + s_1 z + s_2 z^2 + s_3 z^3,
\end{equation}
where the coefficients are defined as $s_0 = 0.0842$, $s_1 = 0.0532$, $s_2 = 0.298$, and $s_3 = -0.0113$. Our findings indicate that the presence of magnification bias could lead to deviations of up to \(6\%\) in the estimates of \(E_G\). This magnitude of bias is significant and should be taken into account for future high-precision estimates of \(E_G\). Consequently, we emphasize that magnification bias must be carefully addressed in subsequent analyses to ensure that the results accurately reflect the underlying gravitational theory.

\section{conclusions}\label{sec:summary}
The \( E_G \) statistic, which integrates gravitational lensing and LSS, represents a valuable cosmological probe for testing theories of gravity, particularly because it is independent of galaxy bias and \( \sigma_8 \). Unlike traditional methods relying on spectroscopic surveys, this study estimates \( E_G \) at four effective redshifts, utilizing photometric redshift data from the DES MagLim sample alongside Planck 2018 CMB lensing convergence maps. To address the significant redshift uncertainties inherent in photometric redshift surveys, we adopt a novel approach for estimating the RSD parameter \( \beta = \hat{f}/\hat{b} \), where \( \hat{f}=f\sigma_8 \) and \( \hat{b}=b\sigma_8 \).

For the growth rate parameter \( \hat{f} \), we compile current measurements from various LSS spectroscopic redshift surveys and implement an ANN algorithm, \texttt{ReFANN}, to derive estimates. The ANN-based predictions for \( \hat{f} \) demonstrate consistency with the standard \( \Lambda \)CDM model at the 68\% confidence level.

For the linear bias parameter \( \hat{b} \), we utilize constraints on \( b(z) \) and \( \sigma_8(z) \) across four redshift bins provided by the DES collaboration to derive corresponding \( \hat{b}(z) \) values. However, the derived value of \( \hat{b} \) for the highest redshift bin (\( z=0.8 \)) is notably low, potentially introducing systematic uncertainties in the final \( E_G \) measurements.

We estimate the \( E_G \) statistic by analyzing the angular power spectra \( C_{\ell}^{\mathrm{gg}} \) and \( C_{\ell}^{\mathrm{g\kappa}} \) . Our results for \( E_G(\ell) \) reveal no significant scale dependence across all redshift bins. Moreover, we present new measurements of the \( E_G \) statistic: \(E_G = 0.354 \pm 0.146\), \(0.452 \pm 0.092\), \(0.414 \pm 0.069\), and \(0.296 \pm 0.069\) (68\% C.L.) for redshifts \( z = 0.30\), \( 0.47\), \( 0.63\), and \( 0.80 \), respectively. These estimates are generally consistent with other \( E_G \) measurements and predictions from \( \Lambda \)CDM, though the statistical uncertainties remain relatively large. Additionally, the \( E_G \) measurement in the fourth redshift bin is notably lower than the theoretical prediction, likely due to an underestimation of the bias in the DES sample at this redshift.

Given the substantial uncertainties in current observational data, we simulate future data from the forthcoming CSST and CMB-S4 experiment to project potential improvements in \( E_G \) precision. Our simulations indicate that future surveys could reduce the uncertainties in \( E_G \) measurements to the 1\% level, enabling a definitive distinction between General Relativity and various modified gravity models.

Lastly, we assess the impact of magnification bias on the \( E_G \) estimates. Our findings suggest that magnification bias could introduce deviations of up to 6\% in the \( E_G \) measurements. This level of bias is significant, and we underscore the importance of carefully accounting for magnification bias in future high-precision \( E_G \) analyses to ensure the accuracy and reliability of constraints on gravitational theories.

\section*{acknowledgments}
This work is supported by the National Natural Science Foundation of China, under grant Nos. 12473004 and 12021003, the National Key R\&D Program of China, No. 2020YFC2201603, the China Manned Space Program through its Space Application System, and the Fundamental Research Funds for the Central Universities.

\bibliography{eg}{}

\begin{thebibliography}{}
\expandafter\ifx\csname natexlab\endcsname\relax\def\natexlab#1{#1}\fi
\providecommand{\url}[1]{\href{#1}{#1}}
\providecommand{\dodoi}[1]{doi:~\href{http://doi.org/#1}{\nolinkurl{#1}}}
\providecommand{\doeprint}[1]{\href{http://ascl.net/#1}{\nolinkurl{http://ascl.net/#1}}}
\providecommand{\doarXiv}[1]{\href{https://arxiv.org/abs/#1}{\nolinkurl{https://arxiv.org/abs/#1}}}

\bibitem[{Abazajian {et~al.}(2022)Abazajian, Abdulghafour, Addison, Adshead, Ahmed, Ajello, Akerib, Allen, Alonso, Alvarez, {et~al.}}]{abazajian2022snowmass}
Abazajian, K., Abdulghafour, A., Addison, G.~E., {et~al.} 2022, arXiv preprint arXiv:2203.08024

\bibitem[{Abazajian {et~al.}(2016)Abazajian, Adshead, Ahmed, Allen, Alonso, Arnold, Baccigalupi, Bartlett, Battaglia, Benson, {et~al.}}]{abazajian2016cmb}
Abazajian, K.~N., Adshead, P., Ahmed, Z., {et~al.} 2016, arXiv preprint arXiv:1610.02743

\bibitem[{Abbott {et~al.}(2023)Abbott, Aguena, Alarcon, Alves, Amon, Andrade-Oliveira, Annis, Avila, Bacon, Baxter, {et~al.}}]{abbott2023dark}
Abbott, T., Aguena, M., Alarcon, A., {et~al.} 2023, Physical Review D, 107, 083504

\bibitem[{Abbott {et~al.}(2022)Abbott, Aguena, Alarcon, Allam, Alves, Amon, Andrade-Oliveira, Annis, Avila, Bacon, {et~al.}}]{abbott2022dark}
Abbott, T.~M., Aguena, M., Alarcon, A., {et~al.} 2022, Physical Review D, 105, 023520

\bibitem[{Aghanim {et~al.}(2020)Aghanim, Akrami, Ashdown, Aumont, Baccigalupi, Ballardini, Banday, Barreiro, Bartolo, Basak, {et~al.}}]{aghanim2020planck}
Aghanim, N., Akrami, Y., Ashdown, M., {et~al.} 2020, Astronomy \& Astrophysics, 641, A8

\bibitem[{Alam {et~al.}(2017{\natexlab{a}})Alam, Miyatake, More, Ho, \& Mandelbaum}]{alam2017testing}
Alam, S., Miyatake, H., More, S., Ho, S., \& Mandelbaum, R. 2017{\natexlab{a}}, Monthly Notices of the Royal Astronomical Society, 465, 4853

\bibitem[{Alam {et~al.}(2017{\natexlab{b}})Alam, Ata, Bailey, Beutler, Bizyaev, Blazek, Bolton, Brownstein, Burden, Chuang, {et~al.}}]{Alam:2016hwk}
Alam, S., Ata, M., Bailey, S., {et~al.} 2017{\natexlab{b}}, Monthly Notices of the Royal Astronomical Society, 470, 2617

\bibitem[{Amon {et~al.}(2018)Amon, Blake, Heymans, Leonard, Asgari, Bilicki, Choi, Erben, Glazebrook, Harnois-Deraps, {et~al.}}]{amon2018kids+}
Amon, A., Blake, C., Heymans, C., {et~al.} 2018, Monthly Notices of the Royal Astronomical Society, 479, 3422

\bibitem[{Beutler {et~al.}(2012)Beutler, Blake, Colless, Jones, Staveley-Smith, Poole, Campbell, Parker, Saunders, \& Watson}]{Beutler:2012px}
Beutler, F., Blake, C., Colless, M., {et~al.} 2012, Monthly Notices of the Royal Astronomical Society, 423, 3430

\bibitem[{Beutler {et~al.}(2017)Beutler, Seo, Saito, Chuang, Cuesta, Eisenstein, Gil-Mar{\'\i}n, Grieb, Hand, Kitaura, {et~al.}}]{Beutler:2016arn}
Beutler, F., Seo, H.-J., Saito, S., {et~al.} 2017, Monthly Notices of the Royal Astronomical Society, 466, 2242

\bibitem[{Blake {et~al.}(2012)Blake, Brough, Colless, Contreras, Couch, Croom, Croton, Davis, Drinkwater, Forster, {et~al.}}]{Blake:2012pj}
Blake, C., Brough, S., Colless, M., {et~al.} 2012, Monthly Notices of the Royal Astronomical Society, 425, 405

\bibitem[{Blake {et~al.}(2013)Blake, Baldry, Bland-Hawthorn, Christodoulou, Colless, Conselice, Driver, Hopkins, Liske, Loveday, {et~al.}}]{Blake:2013nif}
Blake, C., Baldry, I.~K., Bland-Hawthorn, J., {et~al.} 2013, Monthly Notices of the Royal Astronomical Society, 436, 3089

\bibitem[{Blake {et~al.}(2016)Blake, Joudaki, Heymans, Choi, Erben, Harnois-Deraps, Hildebrandt, Joachimi, Nakajima, van Waerbeke, {et~al.}}]{blake2016rcslens}
Blake, C., Joudaki, S., Heymans, C., {et~al.} 2016, Monthly Notices of the Royal Astronomical Society, 456, 2806

\bibitem[{Blake {et~al.}(2020)Blake, Amon, Asgari, Bilicki, Dvornik, Erben, Giblin, Glazebrook, Heymans, Hildebrandt, {et~al.}}]{blake2020testing}
Blake, C., Amon, A., Asgari, M., {et~al.} 2020, Astronomy \& Astrophysics, 642, A158

\bibitem[{Chisari {et~al.}(2019)Chisari, Alonso, Krause, Leonard, Bull, Neveu, Villarreal, Singh, McClintock, Ellison, {et~al.}}]{chisari2019core}
Chisari, N.~E., Alonso, D., Krause, E., {et~al.} 2019, The Astrophysical Journal Supplement Series, 242, 2

\bibitem[{Chuang \& Wang(2013)}]{Chuang:2012qt}
Chuang, C.-H., \& Wang, Y. 2013, Mon. Not. R. Astron. Soc, 435, 255

\bibitem[{Chuang {et~al.}(2016)Chuang, Prada, Pellejero-Ibanez, Beutler, Cuesta, Eisenstein, Escoffier, Ho, Kitaura, Kneib, {et~al.}}]{Chuang:2013wga}
Chuang, C.-H., Prada, F., Pellejero-Ibanez, M., {et~al.} 2016, Monthly Notices of the Royal Astronomical Society, 461, 3781

\bibitem[{Davis {et~al.}(2011)Davis, Nusser, Masters, Springob, Huchra, \& Lemson}]{Davis:2010sw}
Davis, M., Nusser, A., Masters, K.~L., {et~al.} 2011, Monthly Notices of the Royal Astronomical Society, 413, 2906

\bibitem[{De~La~Torre {et~al.}(2013)De~La~Torre, Guzzo, Peacock, Branchini, Iovino, Granett, Abbas, Adami, Arnouts, Bel, {et~al.}}]{delaTorre:2013rpa}
De~La~Torre, S., Guzzo, L., Peacock, J., {et~al.} 2013, Astronomy \& Astrophysics, 557, A54

\bibitem[{De~La~Torre {et~al.}(2017)De~La~Torre, Jullo, Giocoli, Pezzotta, Bel, Granett, Guzzo, Garilli, Scodeggio, Bolzonella, {et~al.}}]{de2017vimos}
De~La~Torre, S., Jullo, E., Giocoli, C., {et~al.} 2017, Astronomy \& Astrophysics, 608, A44

\bibitem[{De~Vicente {et~al.}(2016)De~Vicente, S{\'a}nchez, \& Sevilla-Noarbe}]{de2016dnf}
De~Vicente, J., S{\'a}nchez, E., \& Sevilla-Noarbe, I. 2016, Monthly Notices of the Royal Astronomical Society, 459, 3078

\bibitem[{Dizgah \& Durrer(2016)}]{dizgah2016lensing}
Dizgah, A.~M., \& Durrer, R. 2016, Journal of Cosmology and Astroparticle Physics, 2016, 035

\bibitem[{Dodelson \& Schneider(2013)}]{dodelson2013effect}
Dodelson, S., \& Schneider, M.~D. 2013, Physical Review D—Particles, Fields, Gravitation, and Cosmology, 88, 063537

\bibitem[{Feix {et~al.}(2017)Feix, Branchini, \& Nusser}]{Feix:2016qhh}
Feix, M., Branchini, E., \& Nusser, A. 2017, Monthly Notices of the Royal Astronomical Society, 468, 1420

\bibitem[{Feix {et~al.}(2015)Feix, Nusser, \& Branchini}]{Feix:2015dla}
Feix, M., Nusser, A., \& Branchini, E. 2015, Physical Review Letters, 115, 011301

\bibitem[{Flaugher {et~al.}(2015)Flaugher, Diehl, Honscheid, Abbott, Alvarez, Angstadt, Annis, Antonik, Ballester, Beaufore, {et~al.}}]{flaugher2015dark}
Flaugher, B., Diehl, H., Honscheid, K., {et~al.} 2015, The Astronomical Journal, 150, 150

\bibitem[{Garc{\'\i}a-Garc{\'\i}a {et~al.}(2021)Garc{\'\i}a-Garc{\'\i}a, Ruiz-Zapatero, Alonso, Bellini, Ferreira, Mueller, Nicola, \& Ruiz-Lapuente}]{garcia2021growth}
Garc{\'\i}a-Garc{\'\i}a, C., Ruiz-Zapatero, J., Alonso, D., {et~al.} 2021, Journal of Cosmology and Astroparticle Physics, 2021, 030

\bibitem[{Gelman \& Rubin(1992)}]{gelman1992inference}
Gelman, A., \& Rubin, D.~B. 1992, Statistical science, 7, 457

\bibitem[{Giannantonio {et~al.}(2016)Giannantonio, Fosalba, Cawthon, Omori, Crocce, Elsner, Leistedt, Dodelson, Benoit-L{\'e}vy, Gaztanaga, {et~al.}}]{giannantonio2016cmb}
Giannantonio, T., Fosalba, P., Cawthon, R., {et~al.} 2016, Monthly Notices of the Royal Astronomical Society, 456, 3213

\bibitem[{Gil-Mar{\'\i}n {et~al.}(2016)Gil-Mar{\'\i}n, Percival, Verde, Brownstein, Chuang, Kitaura, Rodr{\'\i}guez-Torres, \& Olmstead}]{Gil-Marin:2016wya}
Gil-Mar{\'\i}n, H., Percival, W.~J., Verde, L., {et~al.} 2016, Monthly Notices of the Royal Astronomical Society, stw2679

\bibitem[{Gil-Mar{\'\i}n {et~al.}(2018)Gil-Mar{\'\i}n, Guy, Zarrouk, Burtin, Chuang, Percival, Ross, Ruggeri, Tojerio, Zhao, {et~al.}}]{Gil-Marin:2018cgo}
Gil-Mar{\'\i}n, H., Guy, J., Zarrouk, P., {et~al.} 2018, Monthly Notices of the Royal Astronomical Society, 477, 1604

\bibitem[{Gong {et~al.}(2019)Gong, Liu, Cao, Chen, Fan, Li, Li, Li, Zhang, \& Zhan}]{gong2019cosmology}
Gong, Y., Liu, X., Cao, Y., {et~al.} 2019, The Astrophysical Journal, 883, 203

\bibitem[{Hartlap {et~al.}(2007)Hartlap, Simon, \& Schneider}]{hartlap2007your}
Hartlap, J., Simon, P., \& Schneider, P. 2007, Astronomy \& Astrophysics, 464, 399

\bibitem[{Hawken {et~al.}(2017)Hawken, Granett, Iovino, Guzzo, Peacock, De~La~Torre, Garilli, Bolzonella, Scodeggio, Abbas, {et~al.}}]{Hawken:2016qcy}
Hawken, A., Granett, B., Iovino, A., {et~al.} 2017, Astronomy \& Astrophysics, 607, A54

\bibitem[{Hivon {et~al.}(2002)Hivon, G{\'o}rski, Netterfield, Crill, Prunet, \& Hansen}]{hivon2002master}
Hivon, E., G{\'o}rski, K.~M., Netterfield, C.~B., {et~al.} 2002, The Astrophysical Journal, 567, 2

\bibitem[{Hou {et~al.}(2018)Hou, S{\'a}nchez, Scoccimarro, Salazar-Albornoz, Burtin, Gil-Mar{\'\i}n, Percival, Ruggeri, Zarrouk, Zhao, {et~al.}}]{Hou:2018yny}
Hou, J., S{\'a}nchez, A.~G., Scoccimarro, R., {et~al.} 2018, Monthly Notices of the Royal Astronomical Society, 480, 2521

\bibitem[{Howlett {et~al.}(2015)Howlett, Ross, Samushia, Percival, \& Manera}]{Howlett:2014opa}
Howlett, C., Ross, A.~J., Samushia, L., Percival, W.~J., \& Manera, M. 2015, Monthly Notices of the Royal Astronomical Society, 449, 848

\bibitem[{Howlett {et~al.}(2017)Howlett, Staveley-Smith, Elahi, Hong, Jarrett, Jones, Koribalski, Macri, Masters, \& Springob}]{Howlett:2017asq}
Howlett, C., Staveley-Smith, L., Elahi, P.~J., {et~al.} 2017, Monthly Notices of the Royal Astronomical Society, 471, 3135

\bibitem[{Hu \& Okamoto(2002)}]{hu2002mass}
Hu, W., \& Okamoto, T. 2002, The Astrophysical Journal, 574, 566

\bibitem[{Hudson \& Turnbull(2012)}]{Hudson:2012gt}
Hudson, M.~J., \& Turnbull, S.~J. 2012, The Astrophysical Journal Letters, 751, L30

\bibitem[{Huterer {et~al.}(2017)Huterer, Shafer, Scolnic, \& Schmidt}]{Huterer:2016uyq}
Huterer, D., Shafer, D.~L., Scolnic, D.~M., \& Schmidt, F. 2017, Journal of Cosmology and Astroparticle Physics, 2017, 015

\bibitem[{Icaza-Lizaola {et~al.}(2020)Icaza-Lizaola, Vargas-Maga{\~n}a, Fromenteau, Alam, Camacho, Gil-Marin, Paviot, Ross, Schneider, Tinker, {et~al.}}]{Icaza-Lizaola:2019zgk}
Icaza-Lizaola, M., Vargas-Maga{\~n}a, M., Fromenteau, S., {et~al.} 2020, Monthly Notices of the Royal Astronomical Society, 492, 4189

\bibitem[{Kazantzidis \& Perivolaropoulos(2018)}]{kazantzidis2018evolution}
Kazantzidis, L., \& Perivolaropoulos, L. 2018, Physical Review D, 97, 103503

\bibitem[{Lepori {et~al.}(2022)Lepori, Tutusaus, Viglione, Bonvin, Camera, Castander, Durrer, Fosalba, Jelic-Cizmek, Kunz, {et~al.}}]{lepori2022euclid}
Lepori, F., Tutusaus, I., Viglione, C., {et~al.} 2022, Astronomy \& Astrophysics, 662, A93

\bibitem[{Lewis(2013)}]{lewis2013efficient}
Lewis, A. 2013, Physical Review D—Particles, Fields, Gravitation, and Cosmology, 87, 103529

\bibitem[{Limber(1953)}]{limber1953analysis}
Limber, D.~N. 1953, The Astrophysical Journal, 117, 134

\bibitem[{Lin {et~al.}(2022)Lin, Gong, Chen, Chan, Fan, \& Zhan}]{lin2022forecast}
Lin, H., Gong, Y., Chen, X., {et~al.} 2022, Monthly Notices of the Royal Astronomical Society, 515, 5743

\bibitem[{Lyke {et~al.}(2020)Lyke, Higley, McLane, Schurhammer, Myers, Ross, Dawson, Chabanier, Martini, Des~Bourboux, {et~al.}}]{lyke2020sloan}
Lyke, B.~W., Higley, A.~N., McLane, J., {et~al.} 2020, The Astrophysical Journal Supplement Series, 250, 8

\bibitem[{Marques {et~al.}(2024)Marques, Madhavacheril, Darwish, Shaikh, Aguena, Alves, Avila, Bacon, Baxter, Bechtol, {et~al.}}]{marques2024cosmological}
Marques, G., Madhavacheril, M., Darwish, O., {et~al.} 2024, Journal of Cosmology and Astroparticle Physics, 2024, 033

\bibitem[{Marques \& Bernui(2020)}]{marques2020tomographic}
Marques, G.~A., \& Bernui, A. 2020, Journal of Cosmology and Astroparticle Physics, 2020, 052

\bibitem[{Mohammad {et~al.}(2018{\natexlab{a}})Mohammad, Granett, Guzzo, Bel, Branchini, de~La~Torre, Moscardini, Peacock, Bolzonella, Garilli, {et~al.}}]{Mohammad:2017lzz}
Mohammad, F., Granett, B.~R., Guzzo, L., {et~al.} 2018{\natexlab{a}}, Astronomy \& Astrophysics, 610, A59

\bibitem[{Mohammad {et~al.}(2018{\natexlab{b}})Mohammad, Bianchi, Percival, de~La~Torre, Guzzo, Granett, Branchini, Bolzonella, Garilli, Scodeggio, {et~al.}}]{Mohammad:2018mdy}
Mohammad, F., Bianchi, D., Percival, W., {et~al.} 2018{\natexlab{b}}, Astronomy \& Astrophysics, 619, A17

\bibitem[{Nadathur {et~al.}(2019)Nadathur, Carter, Percival, Winther, \& Bautista}]{Nadathur:2019mct}
Nadathur, S., Carter, P.~M., Percival, W.~J., Winther, H.~A., \& Bautista, J.~E. 2019, Physical Review D, 100, 023504

\bibitem[{Okumura {et~al.}(2016)Okumura, Hikage, Totani, Tonegawa, Okada, Glazebrook, Blake, Ferreira, More, Taruya, {et~al.}}]{Okumura:2015lvp}
Okumura, T., Hikage, C., Totani, T., {et~al.} 2016, Publications of the Astronomical Society of Japan, 68, 38

\bibitem[{Omori {et~al.}(2019)Omori, Giannantonio, Porredon, Baxter, Chang, Crocce, Fosalba, Alarcon, Banik, Blazek, {et~al.}}]{omori2019dark}
Omori, Y., Giannantonio, T., Porredon, A., {et~al.} 2019, Physical Review D, 100, 043501

\bibitem[{Percival {et~al.}(2014)Percival, Ross, S{\'a}nchez, Samushia, Burden, Crittenden, Cuesta, Magana, Manera, Beutler, {et~al.}}]{percival2014clustering}
Percival, W.~J., Ross, A.~J., S{\'a}nchez, A.~G., {et~al.} 2014, Monthly Notices of the Royal Astronomical Society, 439, 2531

\bibitem[{Perenon {et~al.}(2021)Perenon, Martinelli, Ili{\'c}, Maartens, Lochner, \& Clarkson}]{perenon2021multi}
Perenon, L., Martinelli, M., Ili{\'c}, S., {et~al.} 2021, Physics of the Dark Universe, 34, 100898

\bibitem[{Pezzotta {et~al.}(2017)Pezzotta, de~La~Torre, Bel, Granett, Guzzo, Peacock, Garilli, Scodeggio, Bolzonella, Abbas, {et~al.}}]{Pezzotta:2016gbo}
Pezzotta, A., de~La~Torre, S., Bel, J., {et~al.} 2017, Astronomy \& Astrophysics, 604, A33

\bibitem[{Porredon {et~al.}(2022)Porredon, Crocce, Elvin-Poole, Cawthon, Giannini, De~Vicente, Rosell, Ferrero, Krause, Fang, {et~al.}}]{porredon2022dark}
Porredon, A., Crocce, M., Elvin-Poole, J., {et~al.} 2022, Physical Review D, 106, 103530

\bibitem[{Pullen {et~al.}(2016)Pullen, Alam, He, \& Ho}]{pullen2016constraining}
Pullen, A.~R., Alam, S., He, S., \& Ho, S. 2016, Monthly Notices of the Royal Astronomical Society, 460, 4098

\bibitem[{Pullen {et~al.}(2015)Pullen, Alam, \& Ho}]{pullen2015probing}
Pullen, A.~R., Alam, S., \& Ho, S. 2015, Monthly Notices of the Royal Astronomical Society, 449, 4326

\bibitem[{Qin {et~al.}(2019)Qin, Howlett, \& Staveley-Smith}]{Qin:2019axr}
Qin, F., Howlett, C., \& Staveley-Smith, L. 2019, Monthly Notices of the Royal Astronomical Society, 487, 5235

\bibitem[{Reyes {et~al.}(2010)Reyes, Mandelbaum, Seljak, Baldauf, Gunn, Lombriser, \& Smith}]{reyes2010confirmation}
Reyes, R., Mandelbaum, R., Seljak, U., {et~al.} 2010, Nature, 464, 256

\bibitem[{Rodr{\'\i}guez-Monroy {et~al.}(2022)Rodr{\'\i}guez-Monroy, Weaverdyck, Elvin-Poole, Crocce, Carnero~Rosell, Andrade-Oliveira, Avila, Bechtol, Bernstein, Blazek, {et~al.}}]{rodriguez2022dark}
Rodr{\'\i}guez-Monroy, M., Weaverdyck, N., Elvin-Poole, J., {et~al.} 2022, Monthly Notices of the Royal Astronomical Society, 511, 2665

\bibitem[{Samushia {et~al.}(2012)Samushia, Percival, \& Raccanelli}]{Samushia:2011cs}
Samushia, L., Percival, W.~J., \& Raccanelli, A. 2012, Monthly Notices of the Royal Astronomical Society, 420, 2102

\bibitem[{S{\'a}nchez {et~al.}(2014)S{\'a}nchez, Montesano, Kazin, Aubourg, Beutler, Brinkmann, Brownstein, Cuesta, Dawson, Eisenstein, {et~al.}}]{Sanchez:2013tga}
S{\'a}nchez, A.~G., Montesano, F., Kazin, E.~A., {et~al.} 2014, Monthly Notices of the Royal Astronomical Society, 440, 2692

\bibitem[{S{\'a}nchez {et~al.}(2022)S{\'a}nchez, Omori, Chang, Bleem, Crawford, Drlica-Wagner, Raghunathan, Zacharegkas, Abbott, Aguena, {et~al.}}]{sanchez2022mapping}
S{\'a}nchez, J., Omori, Y., Chang, C., {et~al.} 2022, arXiv preprint arXiv:2210.08633

\bibitem[{Shi {et~al.}(2018)Shi, Yang, Wang, Zhang, Mo, van~den Bosch, Luo, Tweed, Li, Liu, {et~al.}}]{Shi:2017qpr}
Shi, F., Yang, X., Wang, H., {et~al.} 2018, The Astrophysical Journal, 861, 137

\bibitem[{Skara \& Perivolaropoulos(2020)}]{skara2020tension}
Skara, F., \& Perivolaropoulos, L. 2020, Physical Review D, 101, 063521

\bibitem[{Song \& Percival(2009)}]{Song:2008qt}
Song, Y.-S., \& Percival, W.~J. 2009, Journal of Cosmology and Astroparticle Physics, 2009, 004

\bibitem[{Tojeiro {et~al.}(2012)Tojeiro, Percival, Brinkmann, Brownstein, Eisenstein, Manera, Maraston, McBride, Muna, Reid, {et~al.}}]{Tojeiro:2012rp}
Tojeiro, R., Percival, W.~J., Brinkmann, J., {et~al.} 2012, Monthly Notices of the Royal Astronomical Society, 424, 2339

\bibitem[{Torrado \& Lewis(2021)}]{torrado2021cobaya}
Torrado, J., \& Lewis, A. 2021, Journal of Cosmology and Astroparticle Physics, 2021, 057

\bibitem[{Turnbull {et~al.}(2012)Turnbull, Hudson, Feldman, Hicken, Kirshner, \& Watkins}]{Turnbull:2011ty}
Turnbull, S.~J., Hudson, M.~J., Feldman, H.~A., {et~al.} 2012, Monthly Notices of the Royal Astronomical Society, 420, 447

\bibitem[{Wang {et~al.}(2020)Wang, Ma, Li, \& Xia}]{wang2020reconstructing}
Wang, G.-J., Ma, X.-J., Li, S.-Y., \& Xia, J.-Q. 2020, The Astrophysical Journal Supplement Series, 246, 13

\bibitem[{Wang {et~al.}(2018)Wang, Zhao, Chuang, Pellejero-Ibanez, Zhao, Kitaura, \& Rodriguez-Torres}]{Wang:2017wia}
Wang, Y., Zhao, G.-B., Chuang, C.-H., {et~al.} 2018, Monthly Notices of the Royal Astronomical Society, 481, 3160

\bibitem[{Wenzl {et~al.}(2024)Wenzl, Bean, Chen, Farren, Madhavacheril, Marques, Qu, Sehgal, Sherwin, \& Van~Engelen}]{wenzl2024constraining}
Wenzl, L., Bean, R., Chen, S.-F., {et~al.} 2024, Physical Review D, 109, 083540

\bibitem[{Wilson(2016)}]{Wilson:2016ggz}
Wilson, M.~J. 2016, arXiv preprint arXiv:1610.08362

\bibitem[{Yang \& Pullen(2018)}]{yang2018calibrating}
Yang, S., \& Pullen, A.~R. 2018, Monthly Notices of the Royal Astronomical Society, 481, 1441

\bibitem[{Zhang {et~al.}(2007)Zhang, Liguori, Bean, \& Dodelson}]{zhang2007probing}
Zhang, P., Liguori, M., Bean, R., \& Dodelson, S. 2007, Physical Review Letters, 99, 141302

\bibitem[{Zhang {et~al.}(2021)Zhang, Pullen, Alam, Singh, Burtin, Chuang, Hou, Lyke, Myers, Neveux, {et~al.}}]{zhang2021testing}
Zhang, Y., Pullen, A.~R., Alam, S., {et~al.} 2021, Monthly Notices of the Royal Astronomical Society, 501, 1013

\bibitem[{Zhao {et~al.}(2019)Zhao, Wang, Saito, Gil-Mar{\'\i}n, Percival, Wang, Chuang, Ruggeri, Mueller, Zhu, {et~al.}}]{Zhao:2018jxv}
Zhao, G.-B., Wang, Y., Saito, S., {et~al.} 2019, Monthly Notices of the Royal Astronomical Society, 482, 3497

\end{thebibliography}
\bibliographystyle{aasjournal}

\end{document}